\documentclass[12pt]{article}
\usepackage{graphicx}

\begin{document}

\baselineskip=18pt plus 1pt minus 1pt
\begin{center} 

{\large\bf Analytic Description of Critical Point Actinides in a 
Transition from Octupole Deformation to Octupole Vibrations} 

\bigskip

{Dennis Bonatsos$^{\#}$\footnote{e-mail: bonat@inp.demokritos.gr},
D. Lenis$^{\#}$\footnote{e-mail: lenis@inp.demokritos.gr}, 
N. Minkov$^\dagger$\footnote{e-mail: nminkov@inrne.bas.bg},
D. Petrellis$^{\#}$ \footnote{e-mail: petrellis@inp.demokritos.gr},
P. Yotov$^\dagger$ \footnote{e-mail: pyotov@inrne.bas.bg} }
\bigskip

{$^{\#}$ Institute of Nuclear Physics, N.C.S.R.
``Demokritos''}

{GR-15310 Aghia Paraskevi, Attiki, Greece}

{$^\dagger$ Institute for Nuclear Research and Nuclear Energy, Bulgarian
Academy of Sciences }

{72 Tzarigrad Road, BG-1784 Sofia, Bulgaria}

\bigskip

{\bf Abstract}

\end{center}

An analytic collective model in which the relative presence of the 
quadrupole and octupole deformations is determined by a parameter ($\phi_0$),
while axial symmetry is obeyed, is developed. The model [to be called 
the analytic quadrupole octupole axially symmetric model (AQOA)] 
involves an infinite well potential, provides predictions 
for energy and $B(EL)$ ratios which depend only on $\phi_0$, 
draws the border between the regions of octupole deformation and octupole 
vibrations in an essentially parameter-independent way, and describes well 
$^{226}$Th and $^{226}$Ra, for which experimental energy data are shown 
to suggest that they lie close to this border. The similarity of the AQOA 
results with $\phi_0=45^{\rm o}$ for ground state band spectra and $B(E2)$ 
transition rates to the predictions of the X(5) model is pointed out. 
Analytic solutions are also obtained for Davidson potentials of the form 
$\beta^2 +\beta_0^4/\beta^2$, leading to the AQOA spectrum through 
a variational procedure.

\bigskip\bigskip

PACS numbers: 21.60.Ev, 21.60.Fw, 21.10.Re 

Section: Nuclear Structure 

\newpage 

{\bf 1. Introduction} 

Rotational nuclear spectra have long been attributed to quadrupole 
deformations \cite{BM}, while octupole deformations [corresponding to 
reflection asymmetric (pearlike) 
shapes] are supposed to occur in certain regions, most notably in the light 
actinides \cite{Schueler,Rohoz,AB,BN}. The hallmark of octupole deformation is
a negative parity band with levels $L^{\pi}=1^-$, $3^-$, $5^-$, \dots, lying 
close to the ground state band and forming with it a single band with 
$L^{\pi} = 0^+$, $1^-$, $2^+$, $3^-$, $4^+$, \dots, while a negative parity 
band lying systematically higher than the ground state band is a footprint 
of octupole vibrations. The transition from the regime of octupole vibrations 
into the region of octupole deformation has been considered by several 
authors \cite{Nazar,Nazar2,Sheline}. 
A complete algebraic classification of the states occuring 
in the simultaneous presence of the quadrupole and octupole degrees of freedom 
has been provided in terms of the spdf-interacting boson model 
\cite{EIPRL,EINPA}, involving free parameters.  
(It should be noted, however, that an alternative 
interpretation of the low-lying negative parity states in the light actinides 
has been provided in terms of clustering \cite{Daley,Buck,Shneidman}.) 

On the other hand, the transitions from vibrational [{\rm U(5)}] shapes 
to axially symmetric deformed [{\rm SU(3)}] and $\gamma$-unstable deformed 
[{\rm SO(6)}] shapes have been recently 
described in terms of the X(5) \cite{IacX5} and E(5) \cite{IacE5} models 
respectively, which utilize an infinite well potential in the $\beta$ degree 
of freedom, leading to parameter-free (up to overall scale factors) predictions
for spectra and transition probabilities.  

It is the aim of the present work to provide an analytic description of the 
light actinides lying near the border between the regions of octupole 
vibrations and octupole deformation, through the use of a model containing the 
minimum number of free parameters. In this direction, the following steps 
are taken: 

1) Quadrupole and octupole deformations are taken into account on equal 
footing, their relative presence decided by the only free parameter in 
the model, $\phi_0$. 

2) Axial symmetry is assumed, in order to keep the problem tractable. 

3) Symmetrization of the wave functions is carried out as in Ref. \cite{Dzy}, 
involving the irreducible representation (irrep) {\rm A} of the group 
{\rm D}$_2$ for the levels of even parity and the irrep {\rm B}$_1$ of 
the same group for the levels of odd parity. 

4) Separation of variables is achieved in a way analogous to the one used
in the framework of the X(5) model \cite{IacX5}. 

5) An infinite well potential is assumed appropriate for the description 
of the border region, as in the E(5) \cite{IacE5} and X(5) \cite{IacX5}
models. 

The predictions of the model, to be called AQOA, are compared to spectra and 
$B(EL)$ ratios 
for $^{226}$Th and $^{226}$Ra, for which evidence from systematics of 
experimental data is presented, suggesting that they lie close to the 
border between octupole deformation and octupole vibrations. 
This border is found to be drawn by the AQOA model in an essentially 
parameter-independent way.  

In addition, solutions for Davidson potentials \cite{Dav} of the form 
$\beta^2 +\beta_0^4/\beta^2$ are obtained, and a variational method 
\cite{PLB584,PRC70} leading from the Davidson results to the AQOA predictions 
is worked out. 

A different approach to the problem of phase transition in the octupole mode
has been recently given in Ref. 
\cite{Bizzeti}, where the starting point is the introduction of a new 
parametrization of the quadrupole and octupole degrees of freedom, using 
as intrinsic frame of reference the principal axes of the overall tensor 
of inertia, as resulting from the combined quadrupole and octupole deformation.
Comparisons between the results of the two methods are deferred to the
appropriate sections. Three main differences between the two models are:

1) The AQOA model is analytic, while the model of Ref. \cite{Bizzeti} is not.

2) In the AQOA model the quadrupole and octupole degrees of freedom are taken 
into account on equal footing, while in the special form of the model 
of Ref. \cite{Bizzeti} used for comparison to experiment, the octupole 
degree of freedom remains active, while the quadrupole degree of freedom 
is ``frozen'' to a constant value. 

3) In the AQOA model the symmetry axes of the quadrupole and octupole 
deformations are taken to coincide, in order to guarantee axial symmetry, 
while in the more general framework of Ref. \cite{Bizzeti} nonaxial 
contributions, small but not frozen to zero, are taken into account. 

In Section 2 the AQOA model is formulated, while numerical results are 
given in Section 3 and compared to experiment in Section 4. In Section 5 
the variational procedure is described, while Section 6 contains discussion 
of the present results and plans for further work. 

{\bf 2. The Analytic Quadrupole Octupole Axially Symmetric (AQOA) Model} 

{\bf 2.1 Formulation} 

We consider a nucleus in which quadrupole deformation ($\beta_2$) and octupole 
deformation ($\beta_3$) coexist. We take only axially symmetric deformations 
into account, which implies that the $\gamma$ degrees of freedom are ignored,
as in the Davydov--Chaban approach \cite{Chaban}. The body-fixed axes 
$x'$, $y'$, $z'$ are taken along the principal axes of inertia of the 
(axially symmetric) nucleus, while their orientation relative to the 
laboratory-fixed axes $x$, $y$, $z$ is described by the Euler angles 
$\theta = \{\theta_1, \theta_2, \theta_3\}$. 
The Hamiltonian reads \cite{Dzy,Den}
\begin{equation}\label{eq:e1}
H = -\sum_{\lambda=2,3} {\hbar^2 \over 2 B_\lambda} {1\over \beta_\lambda^3} 
{\partial \over \partial \beta_\lambda} \beta_\lambda^3 {\partial \over 
\partial \beta_\lambda} + {\hbar^2 \hat {L^2} \over 6(B_2 \beta_2^2 + 
2 B_3 \beta_3^2) } + V(\beta_2,\beta_3)  
\end{equation}
where $B_2$, $B_3$ are the mass parameters. 

We seek solutions of the Schr\"odinger equation of the form \cite{Dzy}
\begin{equation}\label{eq:e2}
\Phi^{\pm}_L(\beta_2,\beta_3,\theta) = (\beta_2 \beta_3)^{-3/2} 
\Psi^{\pm}_L(\beta_2,\beta_3) \vert LM0,\pm\rangle,
\end{equation} 
where the function $\vert LM0, \pm\rangle$ describes the rotation of an 
axially symmetric nucleus with angular momentum projection $M$ onto the 
laboratory-fixed $z$-axis and projection $K=0$ onto the body-fixed 
$z'$-axis. The moment of inertia with respect to the symmetry axis $z'$ is 
zero, implying that levels with $K \neq 0$ lie infinitely high in energy
\cite{Dzy}. 
Therefore in this model we are restricted to states with $K=0$ only. 
The function $\vert LM0, +\rangle$ transforms according to the irreducible 
representation (irrep) {\rm A} of the group {\rm D}$_2$, while the function 
$\vert LM0,-\rangle$ transforms according to the irrep {\rm B}$_1$ of the same 
group \cite{Dzy,Den}. The general form of these functions is \cite{BM}  
\begin{equation}\label{eq:e3}
\vert LMK, \pm\rangle = \sqrt{ 2L+1 \over 16 \pi^2 (1+\delta_{K0})}
({\cal D}^L_{K,M}(\theta) \pm (-1)^L {\cal D}^L_{-K,M}(\theta)).
\end{equation}
In the special case of $K=0$ it is clear that $\vert LM0,+\rangle\neq 0$ 
for $L=0$, 2, 4, \dots, while $\vert LM0,-\rangle \neq 0$ for $L=1$, 3, 5, 
\dots The functions $\Psi_L^+(\beta_2,\beta_3)$ and 
$\Psi_L^-(\beta_2,\beta_3)$ 
are respectively symmetric and antisymmetric with respect to reflection 
in the plane $x'y'$, and therefore describe states with positive and negative 
parity respectively \cite{Den}. 

Using the solutions of Eq. (\ref{eq:e2}) for the Hamiltonian of Eq. 
(\ref{eq:e1}) the Schr\"odinger equation takes the simplified form 
%\begin{equation}\label{eq:e4} 
$$
\left[ -{\hbar^2\over 2 B_2} {\partial^2 \over \partial \beta_2^2} 
      -{\hbar^2\over 2 B_3} {\partial^2 \over \partial \beta_3^2}
+ {\hbar^2 L(L+1) \over 6 (B_2 \beta_2^2+ 2 B_3 \beta_3^2)} \right.
$$ 
\begin{equation}\label{eq:e4}
\left.
+V(\beta_2,\beta_3) +{3\hbar^2 \over 8} \left({ 1\over B_2 \beta_2^2}
+{1\over B_3 \beta_3^2} \right)-E_L \right] \Psi^\pm_L(\beta_2,\beta_3)=0.   
\end{equation}
This equation is further simplified by introducing \cite{Dzy,Den} 
\begin{equation}\label{eq:e5}
\tilde \beta_2 = \beta_2 \sqrt{B_2\over B}, \quad \tilde \beta_3 = \beta_3 
\sqrt{B_3\over B}, \quad B= {B_2+B_3 \over 2}, 
\end{equation}
as well as reduced energies $\epsilon=(2B/\hbar^2) E$ and reduced potentials 
$u=(2B/\hbar^2) V$ \cite{IacX5,IacE5}, reaching the form 
\begin{equation}\label{eq:e6} 
\left[ -{\partial^2 \over \partial \tilde \beta_2^2} 
       -{\partial^2 \over \partial \tilde \beta_3^2} 
+ {L(L+1)\over 3(\tilde \beta_2^2 + 2 \tilde \beta_3^2)} 
+u(\tilde \beta_2, \tilde \beta_3) + {3\over 4} 
\left( {1\over \tilde \beta_2^2}+{1\over \tilde \beta_3^2}\right)-\epsilon_L
\right] \Psi_L^{\pm}(\tilde \beta_2,\tilde \beta_3) =0.  
\end{equation}
Further simplification occurs through the introduction of polar coordinates 
(with $0\leq \tilde \beta < \infty$ and $-\pi/2 \leq \phi \leq \pi/2$) 
\cite{Dzy,Den} 
\begin{equation}\label{eq:e7} 
\tilde \beta_2 = \tilde \beta \cos \phi, \quad \tilde \beta_3 = \tilde \beta
\sin \phi, \quad \tilde \beta = \sqrt{\tilde \beta_2^2 + \tilde \beta_3^2}, 
\end{equation} 
leading to 
\begin{equation}\label{eq:e8}
\left[ -{\partial^2 \over \partial \tilde \beta^2} -{1\over \tilde \beta} 
{\partial \over \partial \tilde \beta} +{L(L+1) \over 3 \tilde \beta^2 
(1+\sin^2\phi) } -{1\over \tilde \beta^2} {\partial^2 \over \partial \phi^2} 
+ u(\tilde \beta,\phi) + {3\over  \tilde \beta^2 \sin^2 2\phi}-\epsilon_L
\right] \Psi_L^{\pm}(\tilde \beta,\phi) =0. 
\end{equation}
It is clear that $\phi=0$ corresponds to quadrupole deformation alone, 
while $\phi=\pm \pi/2$ corresponds to octupole deformation alone. It is 
worth noticing that the transformation of Eq. (\ref{eq:e7}) allows $\beta_3$ 
to assume both positive and negative values, while $\beta_2$ takes only 
positive values. 

Separation of variables in Eq. (\ref{eq:e8}) can be achieved by assuming 
the potential to be of the form 
$u(\tilde \beta,\phi) = u(\tilde\beta) + u(\tilde \phi^\pm)$, where 
$u(\tilde \phi^\pm )$ 
is supposed 
to be of the form of two very steep harmonic oscillators centered at 
the values $\pm \phi_0$, i.e. 
\begin{equation}\label{eq:e8a}
u (\tilde \phi^\pm )= {1\over 2} c (\phi \mp \phi_0)^2 = {1\over 2} c 
 (\tilde \phi^\pm)^2, \qquad \tilde \phi^\pm = \phi \mp \phi_0,  
\end{equation}
with $c$ being a large constant. In other words,
the nucleus is supposed to be rigid with respect to the variable 
$\phi$, implying that $\phi$ remains close to $\pm \phi_0$ 
and, therefore,
the relative amount of quadrupole and octupole deformation remains constant,
as in Strutinsky-type potential energy calculations \cite{Nazar}. 
This assumption will be (partly) justified {\it a posteriori} by the fact 
that the spectrum remains almost unchanged for values of $\phi_0$ 
between 30$^{\rm o}$ and 60$^{\rm o}$.

In this way Eq. (\ref{eq:e8}) is separated into 
\begin{equation}\label{eq:e9} 
\left[ -{\partial^2 \over \partial \tilde \beta^2} - {1\over \tilde \beta} 
{\partial \over \partial \tilde \beta} +{1\over \tilde\beta^2} \left(
 {L(L+1) \over 3 (1+\sin^2\phi_0)} + {3\over \sin^2 2\phi_0}\right) 
+u(\tilde \beta) -\epsilon_{\tilde\beta}(L) \right] \psi_L^{\pm}(\tilde \beta) 
=0, 
\end{equation}
and 
\begin{equation}\label{eq:e10}
\left[ -{1\over \langle \tilde \beta^2 \rangle}{\partial^2 \over 
\partial(\tilde \phi^\pm)^2} + u (\tilde \phi^\pm ) - \epsilon_{\phi} \right]
 \chi(\tilde \phi^\pm ) =0, 
\end{equation}
where $\Psi_L^{\pm}(\tilde \beta,\phi) = \psi_L^{\pm}(\tilde \beta)
(\chi(\tilde \phi^+) \pm \chi(\tilde \phi^-))/\sqrt{2}$, 
while $\langle \tilde \beta^2\rangle$ is the average of 
$\tilde \beta^2$ over $\psi^{\pm}(\tilde \beta)$, and $\epsilon_L= 
\epsilon_{\tilde \beta}(L) +\epsilon_{\phi}$. 
It is worth noticing that Eq. (\ref{eq:e9}) has the same form for both 
$+\phi_0$ and $-\phi_0$, since only even functions of $\phi_0$ appear in it. 

{\bf 2.2 The $\tilde \beta$-part of the spectrum }

In the case in which $u(\tilde \beta)$ is an infinite well potential 
($u(\tilde\beta)= 0$ if $\tilde\beta \leq \tilde \beta_W$; 
 $u(\tilde \beta)=\infty$ if $\tilde \beta > \tilde \beta_W$),
using the definitions $\epsilon_{\tilde \beta} = k_{\tilde\beta}^2$, 
$z=\tilde \beta k_{\tilde \beta}$, Eq. (\ref{eq:e9}) is brought into the 
form of a Bessel equation 
\begin{equation}\label{eq:e11} 
{d^2\psi_\nu^{\pm} \over dz^2} + {1\over z} {d\psi_\nu^{\pm} \over dz} 
+\left [ 1- {\nu^2 \over z^2} \right] \psi_\nu^{\pm} =0, 
\end{equation}
with 
\begin{equation}\label{eq:e12}
\nu = \sqrt{ {L(L+1)\over 3 (1+\sin^2\phi_0)} + {3\over \sin^2 2\phi_0} } .
\end{equation}
Then the boundary condition $\psi_\nu^{\pm}(\tilde \beta_W)=0$ determines 
the spectrum 
\begin{equation}\label{eq:e13} 
\epsilon_{\tilde \beta, s, \nu} = \epsilon_{\tilde \beta, s, \phi_0, L} 
= (k_{s,\nu})^2, \qquad k_{s,\nu}= {x_{s,\nu} \over \tilde \beta_W}, 
\end{equation}
and the eigenfunctions 
\begin{equation}\label{eq:e14} 
\psi_{s,\nu}^{\pm}(\tilde\beta) = \psi^\pm_{s,\phi_0,L}(\tilde \beta) = 
c_{s,\nu} J_{\nu}(k_{s,\nu} \tilde \beta), 
\end{equation}
where $x_{s,\nu}$ is the $s$th zero of the Bessel function $J_\nu(z)$, 
while $c_{s,\nu}$ are normalization constants, determined from the 
condition $\int_0^{\tilde \beta_W} 
\vert \psi^\pm_{s,\nu}(\tilde \beta)\vert^2 \tilde \beta 
d\tilde \beta =1$ to be $c_{s,\nu}=\sqrt{2}/J_{\nu+1}(k_{s,\nu})$. 
The notation has been kept similar to Ref. \cite{IacX5}. 

Eq. (\ref{eq:e9}) is also exactly soluble \cite{Elliott,Rowe} in the case of 
the Davidson potentials
 \cite{Dav} 
\begin{equation}\label{eq:e15} 
u(\tilde \beta) = \tilde \beta^2 +{\tilde \beta_0^4 \over \tilde \beta^2}.
\end{equation}
In this case the second term of Eq. (\ref{eq:e15}) is combined with the third 
term of Eq. (\ref{eq:e9}), leading to eigenfunctions which are Laguerre 
polynomials 
\begin{equation}\label{eq:e16}
F^L_n(\tilde \beta) = \sqrt{2n! \over \Gamma( n+a+1)} \tilde\beta^{a} 
L_n^a(\tilde \beta^2) e^{-\tilde \beta^2/2}, 
\end{equation}
where 
\begin{equation}\label{eq:e17}
a= \sqrt{ {L(L+1)\over 3(1+\sin^2\phi_0)} + {3\over \sin^2 2\phi_0} +\beta_0^4}
\end{equation}
while the energy eigenvalues are given by 
\begin{equation}\label{eq:e18}
E_{n,L} = 2n+a +1 = 2n+1 + \sqrt{ {L(L+1)\over 3(1+\sin^2 \phi_0)} + {3\over 
\sin^2 2\phi_0} +\beta_0^4} . 
\end{equation}
It is worth remarking that the excitation energies, $E_{0,L}-E_{0,0}$, 
within the ground state band (which is characterized by $n=0$), divided 
by an appropriate normalization constant read 
\begin{equation}\label{eq:e19}
E'_{0,L,exc} = \sqrt{ 1 + b_H L(L+1)} -1 , 
\end{equation}
with 
$ b_H^{-1} = 3(1+\sin^2 \phi_0) \left( {3\over \sin^2 2 \phi_0} +\beta_0^4
\right) $. 
Eq. (\ref{eq:e19}) is the Holmberg--Lipas formula \cite{Lipas}. 

In what follows, the infinite well potential will be used everywhere.
Davidson potentials will be briefly employed in Section 5. 

{\bf 2.3 The $\phi$-part of the spectrum} 

Eq. (\ref{eq:e10}) for the potential of Eq. (\ref{eq:e8a}) takes the form 
\begin{equation}\label{eq:e20}
\left[ -{\partial^2 \over \partial (\tilde \phi^\pm)^2} +{1\over 2} c \langle 
\tilde \beta^2 \rangle (\tilde \phi^\pm)^2 \right] \chi(\tilde \phi^\pm)
= \epsilon_\phi \langle \tilde \beta^2 \rangle \chi(\tilde\phi^\pm),
\end{equation}
where $\tilde \phi^\pm = \phi \mp \phi_0$. 
This is a simple harmonic oscillator equation with energy eigenvalues 
\begin{equation}\label{eq:e21}
\epsilon_{\phi} = \sqrt{ 2c\over \langle \tilde \beta^2\rangle } \left(
n_\phi +{1\over 2} \right), \qquad n_\phi = 0,1,2,\ldots 
\end{equation}
and eigenfunctions 
\begin{equation}\label{eq:e22}
\chi_{n_\phi}(\tilde \phi^\pm) = N_{n_\phi} H_{n_\phi} (b \tilde \phi^\pm) 
e^{-b^2 (\tilde \phi^\pm)^2 /2} , \qquad b=\left( c \langle \tilde \beta^2 
\rangle \over 2 \right)^{1/4},
\end{equation}
with normalization constant $ N_{n_\phi} = \sqrt{b\over \sqrt{\pi} 2^{n_\phi}
n_\phi!}$. 

The total energy in the present model is then 
\begin{equation}\label{eq:e24}
E(s,L,\phi_0, n_\phi) = E_0 + A \epsilon_{\tilde \beta,s,\phi_0,L} + B n_\phi.
\end{equation}

{\bf  2.4 $B(EL)$ transition rates}

In the axial case used here the electric quadrupole and octupole operators are 
\begin{equation}\label{eq:e31}
T^{(E2)}_\mu = t_2 \beta_2 {\cal D}^{(2)}_{\mu,0}(\theta), \qquad 
T^{(E3)}_\mu =t_3 \beta_3 {\cal D}^{(3)}_{\mu,0}(\theta), 
\end{equation}
while the electric dipole operator reads \cite{Dzy}
\begin{equation}\label{eq:e32}
T^{(E1)}_\mu = t_1 \beta_2 \beta_3 {\cal D}^{(1)}_{\mu,0}(\theta).
\end{equation}
The total wave function in the case of the infinite well potential is 
\begin{equation}\label{eq:e33}
\Phi^{\pm}_L (\beta_2, \beta_3, \theta) = C (\beta_2 \beta_3)^{-3/2} 
J_\nu(k_{s,\nu} \tilde \beta) {(\chi_{n_\phi}(\tilde \phi^+)\pm 
                                \chi_{n_\phi}(\tilde \phi^-))\over \sqrt{2}} 
\sqrt{2L+1\over 32 \pi^2} (1\pm (-1)^L) {\cal D}^L_{0,M}(\theta), 
\end{equation} 
where $C$ is a constant, 
while in the case of the Davidson potentials the same expression holds 
with $J_\nu(k_{s,\nu} \tilde \beta)$ replaced by $F^L_n(\tilde \beta)$. 

$B(EL)$ transition rates are given by 
\begin{equation}\label{eq:e34} 
B(EL; L_i a_i \to L_f a_f) = { \vert \langle L_f a_f \vert \vert T^{(EL)}
\vert \vert L_i a_i \rangle \vert^2 \over (2L_i+1)}, 
\end{equation}
where the reduced matrix element is obtained through the Wigner-Eckart theorem
\begin{equation}\label{eq:e35} 
\langle L_f \mu_f a_f \vert T^{(EL)}_\mu \vert L_i \mu_i a_i\rangle = 
{ (L_i L L_f \vert \mu_i \mu \mu_f)\over \sqrt{2L_f+1} } \langle L_f a_f 
\vert \vert T^{EL} \vert \vert L_i a_i \rangle . 
\end{equation}

In Eq. (\ref{eq:e34}) the integration over the angles $\theta$ 
involves a standard integral over three Wigner functions \cite{Edmonds}, 
which leads to $(L_i L L_f \vert 000)$, while the rest of the integrations 
are performed over 
$\int\int \beta_2^3 d\beta_2 \beta_3^3 d\beta_3$, where the $\beta_2^3$, 
$\beta_3^3$ factors come from the volume element and cancel with the 
first factor of Eq. (\ref{eq:e33}). Using Eqs. (\ref{eq:e5}) and 
(\ref{eq:e7}), as well as the relevant Jacobian, one finds (up to constant 
factors) that the integration is over $\int \tilde \beta d\tilde \beta d\phi$. 

In the integrals over $\phi$, only the case of $n_\phi =0$, corresponding 
to $H_0=1$, is considered. The results are factors depending 
on the parameters $b$ and $\phi_0$, as well as on the multipolarity 
of the transition. Therefore in Section 3, ratios of $B(EL)$ transition 
rates will be presented, in which these factors cancel out. 

The integrals over $\tilde \beta$ are
\begin{equation}\label{eq:e39}
I_{\tilde \beta}^{(E2)} = I_{\tilde \beta}^{(E3)} 
=\int  \tilde \beta^2  J_{\nu_i}(k_{s_i,\nu_i}\tilde 
\beta) J_{\nu_f}(k_{s_f,\nu_f}\tilde \beta) d\tilde \beta , 
\end{equation}  
\begin{equation}\label{eq:e40}
I_{\tilde \beta}^{(E1)}=\int  \tilde \beta^3  J_{\nu_i}(k_{s_i,\nu_i}\tilde 
\beta) J_{\nu_f}(k_{s_f,\nu_f}\tilde \beta) d\tilde \beta , 
\end{equation}  
in the case of the infinite well potential, while for the Davidson potentials 
the Bessel functions are replaced by Laguerre polynomials, as above. 
The final result then reads 
\begin{equation}\label{eq:e41}
B(EL; L_i \to L_f) = c (c_{s_i,\nu_i} c_{s_f,\nu_f})^2 
(L_i L L_f\vert 000)^2 (I_{\tilde\beta}^{(EL)})^2, 
\end{equation}
where $L=1,2,3$ and all constant factors have been absorbed in $c$. 

{\bf 3. Numerical results} 

Spectra for the ground state band and the negative parity band associated 
with it ($s=1$), as well as for the first excited 
band ($s=2$) and the second excited band ($s=3$), normalized to the $2_1^+$ 
state of the ground state band, are shown for several values of $\phi_0$ 
in Table 1. A few $R(L) = E(L)/E(2)$ ratios are also depicted as functions 
of $\phi_0$ in Fig. 1(a). It is clear that the results are quite stable in the 
region $30^{\rm o} \leq \phi_0 \leq 60^{\rm o}$, while at the limiting cases 
near $\phi_0= 0^{\rm o}$ and $90^{\rm o}$ the rigid rotor results are obtained,
corresponding to a pure rotational spectrum for the ground state band and 
the associated negative parity band, while the excited bands are pushed 
to infinity.   

$B(E2)$ transition rates are listed in Table 2 for several values 
of $\phi_0$, while a few $B(E2)$ ratios are shown in Fig. 1(b) as 
functions of $\phi_0$, their behavior being quite smooth in the region 
$30^{\rm o} \leq \phi_0 \leq 60^{\rm o}$. The same remark applies to 
$B(E1)$ and $B(E3)$ transitions, listed in Tables 3 and 4, and 
shown in Fig. 1(c). 

It is worth remarking that the minima of energy ratios related to the 
ground state band, as well as the maxima of $B(EL)$ ratios regarding 
the ground state band and the associated negative parity band, 
reported in the caption of Fig. 1, are all located between $\phi_0=40^{\rm o}$ 
and $43^{\rm o}$, while the minima of energy ratios regarding the excited 
($s=2$, 3) bands are located near $\phi_0=35^{\rm o}$.  

The tables and figures mentioned so far indicate that the region of interest
in the present model, in which smooth and essentially parameter independent 
behavior of spectra and $B(EL)$ rates is observed, 
is the region $30^{\rm o} \leq \phi_0 \leq 60^{\rm o}$, 
to which further considerations will be limited. 

In addition to the results of the AQOA model, the X(5) spectrum is included 
in Table 1 for comparison. It is clear that the ground state band of X(5) 
lies a little lower than the ground state band of the AQOA model with 
$\phi_0=45^{\rm o}$, while for the $s=2$ and $s=3$ bands the AQOA model 
predictions for $\phi_0=45^{\rm o}$ are larger than the X(5) values by almost 
a factor of two. Furthermore, in Table 2 the $B(E2)$ transitions within 
the ground state band
of X(5) are shown for comparison. It is clear that the X(5) values are 
slightly higher than the corresponding predictions of the AQOA model for 
$\phi_0=45^{\rm o}$.   

The similarities between the ground state bands of the AQOA and X(5) models
can be understood as due to the fact that both models originate from
the Bohr Hamiltonian and use an infinite well potential, while in addition 
for the properties of the ground state band the quadrupole degree of 
freedom, included in both models, is expected to be important. 
In contrast, the excited bands appear to be more sensitive to the 
inclusion of the octupole degree of freedom. The position of the $0_2^+$
state becomes therefore an important factor in the process of comparison to 
experiment. One can also think of the AQOA model as an extension of the X(5)
framework, in which the negative parity states, as well as the $B(EL)$ 
transitions involving them, are included. 

{\bf 4. Comparison to experiment} 

Experimental data for the ground state and related negative parity bands 
of $^{220-234}$Th are shown in Fig. 2(a). It is clear that $^{226}$Th lies 
on the border between two different regions. Below $^{226}$Th the odd--even
staggering is very small, while from $^{228}$Th up the odd--even staggering is 
becoming much larger, increasing with the neutron number $N$. A quantitative 
measure of the odd--even staggering and related figures can be found in 
Ref. \cite{PRC62}. It is clear that below $^{226}$Th the situation corresponds 
to octupole deformation, in which the ground state band and the negative 
parity band merge into a single band, while above $^{226}$Th the picture is 
corresponding to octupole vibrations, i.e. the negative parity band is a 
rotational band built on an octupole bandhead, thus lying systematically 
higher than the ground state band. Theoretical predictions for
$\phi=45^{\rm o}$ lie a little below $^{226}$Th, while the $\phi=60^{\rm o}$ 
results follow the $^{226}$Th data very closely. It is worth remarking that 
the procedure of Ref. \cite{Bizzeti}, which is quite different from the 
present one, also leads to the identification of $^{226}$Th as the 
nucleus lying closest to the transition point from octupole deformation 
to octupole vibrations. 

A similar picture is observed in $^{218-228}$Ra, shown in Fig. 2(b). 
In this case octupole deformation appears below $^{226}$Ra, while $^{228}$Ra 
is already in the regime of octupole vibrations. Theoretical predictions 
for $\phi_0=45^{\rm o}$ again lie a little below $^{226}$Ra, while 
the $^{226}$Ra data are followed quite closely by the predictions of 
$\phi_0=56^{\rm o}$. 

The behavior observed in Fig. 2(a) can be better understood by considering 
Figs. 3(a) and 3(b), where the experimental energy levels of the ground 
state band and the associated octupole band are shown, for the same thorium 
isotopes. While the even parity levels, shown in Fig. 3(a), smoothly 
decrease with increasing neutron number $N$, as a result of increasing 
quadrupole collectivity, the odd parity levels, shown in Fig. 3(b), exhibit 
a minimum, which is located at $N=136$ up to $L=9$, while it moves to $N=138$ 
for higher $L$. This change of behavior is then attributed to the octupole 
degree of freedom, showing that $^{226}_{90}$Th$_{136}$ lies near the border 
between octupole deformation and octupole vibrations.   
The change of behavior is not abrupt, since the effect due to octupole 
deformation is ``moderated'' by the quadrupole deformation 
setting in in parallel. 

In a similar manner the behavior observed in Fig. 2(b) can be clarified by 
considering Figs. 3(c) and 3(d), where the experimental data for the same 
radium isotopes are presented. Again, the even parity levels decrease 
with increasing $N$, while the odd parity levels exhibit a minimum, located 
at $N=136$ up to $L=5$, while it moves to $N=138$ for higher $L$, showing 
that $^{226}_{88}$Ra$_{138}$ lies close to the border between the regions 
of octupole deformation and octupole vibrations. 

The transition from octupole deformation to octupole vibrations can also be 
seen by considering the simplest quantity measuring the relative displacement 
of the negative parity levels with respect to the even parity ones,
\begin{equation}\label{eq:e51}
\Delta E(L) = E(L)-{E(L-1) + E(L+1) \over 2}. 
\end {equation}
Results for the Th and Ra isotopes are shown  in Figs. 4(a) and 4(b), 
respectively. In Fig. 4(a) it is clear that in $^{222-226}$Th the staggering 
is decreasing rapidly with increasing angular momentum, reaching a vanishing 
value and staying close to it, which is the hallmark of octupole deformation
\cite{Schueler,Phillips},
while in $^{228-234}$Th the decrease is much 
slower and vanishing values, if any, correspond to very high angular momenta, 
a behavior expected for octupole vibrations. 
Again $^{226}$Th appears closest to the border between the two regions.
In Fig. 4(b), $^{220-226}$Ra exhibit the rapid decrease of staggering and 
the sticking to values close to zero beyond the first vanishing value, 
while $^{228-230}$Ra follow the slow decrease pattern. As a result, 
$^{226}$Ra appears to be closest to the border line between the two regions. 

As far as the $0_2^+$ bandhead is concerned, the experimental values 
(normalized to the $2_1^+$ state) are 12.186 for $^{226}$Ra and 11.152 
for $^{226}$Th, in good agreement with the 11.226 and 12.410 values 
predicted by the AQOA model for the $\phi_0$ values of $56^{\rm o}$ and 
$60^{\rm o}$ used in Fig. 2~. (The model of Ref. \cite{Bizzeti} provides 
a value of 8.528 for $^{226}$Th.)
It should be noticed that the normalized $0_2^+$ bandhead 
is lying close to this height for all Ra and Th isotopes for which data
exist, namely $^{222}$Ra (8.225), $^{224}$Ra (10.861), $^{228}$Ra (11.300), 
$^{228}$Th (14.402), $^{230}$Th (11.934), $^{232}$Th (14.794), $^{234}$Th 
(16.347), with data taken from the references used in Fig. 2~. 

Considering the AQOA model as an extension of the X(5) framework involving 
negative parity states, as remarked at the end of Section 3, implies that 
the search for X(5)-like nuclei in the light actinides, where the presence 
of low-lying negative parity bands is important, should be focused 
on nuclei with $R(4)$ ratio close to 3.0 and $0_2^+$ bandhead higher 
than the X(5) value of 5.65~.  

Detailed comparisons to $B(EL)$ transition rates are not feasible, because 
of lack of experimental data. We therefore use ratios of $B(EL)$ transitions, 
also used in earlier work \cite{Bizzeti,Raduta}. Thus in Table 5 and 
Fig. 5(a) the experimental $B(E1;L\to L-1)/B(E2;L\to L-2)$ ratios used 
in Ref. \cite{Bizzeti} are shown, together with theoretical predictions 
from the same source, and predictions for $\phi=45^{\rm o}$ and $60^{\rm o}$,
the values also used in Fig. 2(a). The present theoretical predictions for the 
two different values of $\phi_0$ practically coincide (indicating that the 
predictions are essentially parameter free)  and are in most cases 
within the error bars of the experimental points, while the predictions 
of Ref. \cite{Bizzeti} grow a little faster as a function of angular 
momentum. 

Furthermore, in Table 6 and Fig. 5(b) the experimental 
$B(E1;L\to L+1) / B(E1; L\to L-1)$ ratios \cite{Woll} used in Ref. 
\cite{Raduta} are shown, together with three sets of theoretical predictions 
in the framework of the extended coherent states model (ECSM) 
\cite{Raduta2} from the same source, 
corresponding to the lowest order choice for the $E1$ transition operator 
(R-h), as well as to two different choices of the $E1$ transition operator,
including anharmonic terms assumed suitable for the transition region 
(R-I, R-II) \cite{Raduta}. In addition, predictions for $\phi_0=45^{\rm o}$ 
and $56^{\rm o}$, the same values used in Fig. 2(b), are shown. It is clear 
that the predictions for the two different values of $\phi_0$ practically 
coincide (indicating that the predictions are essentially parameter free)
and in all cases are within the error bars of the experimental points, 
being in very close agreement to the R-I predictions of Ref. \cite{Raduta}. 

On the results presented in this section, the following additional 
comments apply. 

1) Figs. 2 and 5 indicate that $^{226}$Th ($^{226}$Ra) can be well described 
using the AQOA model with $\phi_0=60^{\rm o}$ ($\phi_0=56^{\rm o}$), which 
provides results quite similar to the $\phi_0=45^{\rm o}$ case. In all 
these cases, Eq. (\ref{eq:e7}) [together with Eq. (\ref{eq:e5})]
indicates that the quadrupole and octupole 
deformations are present in comparable amounts. This is in agreement with 
Strutinsky-type potential-energy calculations \cite{Nazar,Leander},
resulting in comparable $\beta_2$ and $\beta_3$ values for these nuclei.  
The presence of octupole deformation in $^{226}$Ra has also been realized 
in a study \cite{Zamfir} within the framework of the spdf-IBM 
\cite{EIPRL,EINPA}. 

2) Figs. 2-4 suggest that $^{226}$Th and $^{226}$Ra lie close to the border 
between octupole deformation and octupole vibrations. This is in agreement 
with Woods--Saxon--Bogolyubov cranking calculations \cite{Nazar2} for the Ra 
and Th isotopes, suggesting shape changes from nearly spherical ($N\simeq 130$)
to octupole-deformed ($N\simeq 134$) to well-deformed reflection-symmetric 
($N\simeq 140$) shapes, in which negative-parity bands can be interpreted 
in terms of octupole vibrations. 

3) One can easily see that no odd--even staggering is predicted by the AQOA 
model. This is in agreement to the well known fact that odd-even staggering 
is produced when the potential in $\beta_3$ is a double well with two 
symmetric minima \cite{Leander2}, the staggering being sensitive to the 
angular momentum dependence of the height of the potential barrier 
\cite{Jolos}. An infinitely high barrier leads to no odd-even staggering 
\cite{Leander2}, which is indeed the case here. The introduction of a finite 
barrier in the present model will lead to staggering, but it will require 
the addition of at least one new parameter, in contrast to the main goal 
of the present work, which is the description of the border between 
octupole deformation and octupole vibrations with the minimum number of 
parameters possible. As shown in Figs. 2(a) and 2(b), the model does 
predict the border between the regions of octupole deformation and octupole
vibrations in an essentially parameter independent way. 

4) It should be noticed that the transition examined here is the one 
from octupole deformation to octupole vibrations as a function of the 
neutron number in a chain of isotopes, which is different from the gradual 
setting in of octupole deformation as a function of angular momentum
in a given nucleus, usually studied by considering the odd-even staggering
\cite{Schueler,Phillips}, as already discussed in relation to Fig. 4. 

{\bf 5. The variational procedure} 

In Refs. \cite{PLB584,PRC70} a variational procedure has been introduced, 
leading from the results of one-parameter Davidson potentials to the 
parameter-free E(5) and X(5) predictions. The same procedure can be applied 
in the present case, by considering (for given $\phi_0$) the $R(L)=E(L)/E(2)$ 
ratios predicted by the Davidson potentials of Eq. (\ref{eq:e15}) for the 
excitation energies of the ground state band and the associated negative 
parity band, and determining for each value of $L$ separately the value 
of the parameter $\beta_0$ at which the derivative of the ratio $R(L)$ 
with respect to $\beta_0$ has a sharp maximum. The collection of $R(L)$ 
values selected in this way (for the case of $\phi=45^{\rm o}$) 
is shown in Table 7 and Fig. 6, together with the limiting cases of 
$\beta_0=0$ (a vibrator) and $\beta_0\to \infty$ (a rigid rotor). 
It is clear that the collection created through the variational 
procedure practicaly coincides with the predictions of the present 
model utilizing an infinite well potential, thus indicating that the choice 
of the infinite well potential indeed correponds to the transition point 
between a vibrator ($\beta_0=0$) and a rigid rotor ($\beta_0\to \infty$), 
since it is at the transition point that the rate of change of the $R(L)$ 
ratios is expected to become maximum.  

{\bf 6. Discussion} 

The analytic quadrupole octupole axially symmetric (AQOA) model introduced 
in this 
work describes well the border between octupole deformation and octupole 
vibrations in the light actinides, which corresponds to $^{226}$Th and 
$^{226}$Ra in the Th and Ra isotopic chains respectively. Some of the main 
ingredients of the present model, such as the infinite well potential and the 
approximate separation of variables, strongly resemble the ones used 
in the X(5) model, describing the critical point of the shape phase 
transition from vibrational to axially deformed rotational nuclei 
\cite{IacX5}, determined through the study of potential energy surfaces 
derived from the Hamiltonian of the Interacting Boson Model \cite{IA}. 
An interesting task is the study of the potential energy surfaces 
resulting in the spdf-IBM \cite{EIPRL,EINPA}, the version of IBM including 
the octupole degree of freedom in addition to the quadrupole one, 
which can possibly lead to the determination of a shape phase transition from 
octupole deformation to octupole vibrations, in a manner similar to the 
determination of the critical point between the spherical and triaxial 
shapes found recently through the study of the potential energy surfaces 
resulting from an IBM-2 Hamiltonian \cite{Arias,Caprio}.  
Although some early results 
are given in Ref. \cite{EINPA}, this task is far from complete.
The persistence of axial symmetry, as well as the importance of parity 
projection in this context have been emphasized \cite{Kuyucak,Honma}.  
The inclusion of staggering in the present model, as well as its application
to the rare earth region near $A=150$, where octupole deformation is known 
to occur \cite{AB,BN}, are also of interest. 

{\bf Acknowledgements} 

Enlightening discussions with Professor F. Iachello  are gratefully 
acknowledged.

\newpage 
\parindent=0pt
%%%%%%%%%%%%%%%%%%%%%%%%%%%%%%%%%%%%%%%%%%%%%%%%%%%%%%%%%%%%%%%%%%%%%%
%%%%%%%%%%%%%%%%%%% Table 1 %%%%%%%%%%%%%%%%%%%%%%%%%%%%%%%%%%%%%%%%

\begin{table}

\caption{Spectra of the AQOA model for the ground state band and the 
associated negative 
parity band ($s=1$), as well as for the first excited band ($s=2$) and the 
second excited band ($s=3$), normalized to the energy of 
the $2_1^+$ state, for different values of $\phi_0$.
The second column contains the values obtained slightly above $0^{\rm o}$
or slightly below $90^{\rm o}$. In addition, the X(5) spectrum is shown for 
comparison. See Section 3 for further discussion.} 

\bigskip

\begin{tabular}{r r r r r r r r}
\hline
$\phi_0$ & $0^{\rm o}$, $90^{\rm o}$ & $15^{\rm o}$ & $30^{\rm o}$ & 
$45^{\rm o}$ & $60^{\rm 0}$ & $75^{\rm o}$ & X(5) \\ 
$L^{\pi}$ &  &  &  &  &  &  & \\
 \hline
$s=1$ &    &       &       &       &       &   &    \\
$0^+$  & 0.000 & 0.000 & 0.000 & 0.000 & 0.000 & 0.000 & 0.000 \\
$1^-$  & 0.333 & 0.337 & 0.344 & 0.346 & 0.342 & 0.336 &       \\
$2^+$  & 1.000 & 1.000 & 1.000 & 1.000 & 1.000 & 1.000 & 1.000 \\
$3^-$  & 2.000 & 1.969 & 1.921 & 1.912 & 1.938 & 1.981 &       \\
$4^+$  & 3.333 & 3.221 & 3.069 & 3.039 & 3.119 & 3.264 & 2.904 \\
$5^-$  & 5.000 & 4.734 & 4.414 & 4.351 & 4.513 & 4.832 &       \\
$6^+$  & 7.000 & 6.490 & 5.935 & 5.829 & 6.098 & 6.667 & 5.430 \\
$7^-$  & 9.333 & 8.471 & 7.620 & 7.459 & 7.857 & 8.755 &       \\ 
$8^+$  &12.000 &10.666 & 9.459 & 9.233 & 9.779 &11.082 & 8.483 \\
$9^-$  &15.000 &13.065 &11.445 &11.144 &11.857 &13.635 &       \\
$10^+$ &18.333 &15.659 &13.574 &13.187 &14.082 &16.406 &12.027 \\
$11^-$ &22.000 &18.443 &15.841 &15.359 &16.451 &19.386 &       \\
$12^+$ &26.000 &21.410 &18.245 &17.658 &18.959 &22.567 &16.041 \\
$13^-$ &30.333 &24.557 &20.782 &20.081 &21.605 &25.943 &       \\
$14^+$ &35.000 &27.881 &23.452 &22.626 &24.384 &29.510 &20.514 \\
$15^-$ &40.000 &31.379 &26.251 &25.293 &27.297 &33.264 &       \\
$16^+$ &45.333 &35.048 &29.180 &28.080 &30.340 &37.200 &25.437 \\
$17^-$ &51.000 &38.886 &32.237 &30.985 &33.513 &41.315 &       \\
$18^+$ &57.000 &42.892 &35.421 &34.009 &36.814 &45.607 &30.804 \\
$19^-$ &63.333 &47.064 &38.731 &37.150 &40.242 &50.074 &       \\
$20^+$ &70.000 &51.402 &42.166 &40.408 &43.796 &54.713 &36.611 \\
\hline
$s=2$ &    &       &       &       &       &    &   \\
$0^+$  &       &13.292 & 8.983 & 9.351 &12.410 &23.896 & 5.649 \\
$2^+$  &       &14.893 &10.726 &11.133 &14.160 &25.502 & 7.450 \\
$4^+$  &       &18.384 &14.204 &14.630 &17.763 &29.098 &10.689 \\
$6^+$  &       &23.392 &18.820 &19.209 &22.649 &34.410 &14.751 \\
\hline
$s=3$ &    &       &       &       &       &     &   \\
$0^+$  &       &30.940 &21.944 &23.114 &30.316 &55.625 &14.119 \\
\hline 
\end{tabular}
\end{table}

\newpage 
\parindent=0pt
%%%%%%%%%%%%%%%%%%%%%%%%%%%%%%%%%%%%%%%%%%%%%%%%%%%%%%%%%%%%%%%%%%%%%%
%%%%%%%%%%%%%%%%%%% Table 2  %%%%%%%%%%%%%%%%%%%%%%%%%%%%%%%%%%%%%%%%

\begin{table}

\caption{ $B(E2; L_i \to L_f)$ values between states of the AQOA model with 
$s=1$. $B(E2)$s with $L_i$ and $L_f$ even are normalized to the 
$2_1^+\to 0_1^+$ transition, while $B(E2)$s with $L_i$ and $L_f$ odd 
are normalized to the 
$3_1^-\to 1_1^-$ transition. The X(5) results are also shown for comparison. 
See Section 3 for further discussion.} 

\bigskip

\begin{tabular}{r r r r r r r r r }
\hline
   & $\phi_0$ & $1^{\rm o}$ & $15^{\rm o}$ & $30^{\rm o}$ & 
$45^{\rm o}$ & $60^{\rm 0}$ & $75^{\rm o}$ & X(5) \\ 
$L_i^{\pi}$ & $L_f^{\pi}$ &  &  &  &  &  & &  \\
 \hline
$ 2^+$ &$ 0^+$ & 1.000 & 1.000 & 1.000 & 1.000 & 1.000 & 1.000 & 1.000 \\
$ 4^+$ &$ 2^+$ & 1.429 & 1.475 & 1.530 & 1.539 & 1.509 & 1.457 & 1.599 \\
$ 6^+$ &$ 4^+$ & 1.574 & 1.701 & 1.834 & 1.855 & 1.786 & 1.656 & 1.982 \\
$ 8^+$ &$ 6^+$ & 1.648 & 1.870 & 2.072 & 2.104 & 2.005 & 1.797 & 2.276 \\
$10^+$ &$ 8^+$ & 1.693 & 2.011 & 2.268 & 2.309 & 2.187 & 1.915 & 2.509 \\
$12^+$ &$10^+$ & 1.723 & 2.131 & 2.431 & 2.480 & 2.342 & 2.019 & 2.697 \\
$14^+$ &$12^+$ & 1.746 & 2.236 & 2.569 & 2.626 & 2.476 & 2.111 & 2.854 \\
$16^+$ &$14^+$ & 1.762 & 2.327 & 2.687 & 2.751 & 2.593 & 2.194 & 2.987 \\
$18^+$ &$16^+$ & 1.776 & 2.407 & 2.790 & 2.860 & 2.695 & 2.269 & 3.101 \\
$20^+$ &$18^+$ & 1.787 & 2.478 & 2.881 & 2.955 & 2.785 & 2.337 & 3.200 \\
\hline
$ 3^-$ &$ 1^-$ & 1.000 & 1.000 & 1.000 & 1.000 & 1.000 & 1.000 &       \\
$ 5^-$ &$ 3^-$ & 1.179 & 1.227 & 1.278 & 1.286 & 1.260 & 1.210 &       \\
$ 7^-$ &$ 5^-$ & 1.257 & 1.374 & 1.479 & 1.496 & 1.444 & 1.335 &       \\
$ 9^-$ &$ 7^-$ & 1.301 & 1.491 & 1.642 & 1.665 & 1.595 & 1.433 &       \\
$11^-$ &$ 9^-$ & 1.329 & 1.591 & 1.777 & 1.806 & 1.723 & 1.518 &       \\
$13^-$ &$11^-$ & 1.350 & 1.677 & 1.890 & 1.924 & 1.832 & 1.594 &       \\
$15^-$ &$13^-$ & 1.365 & 1.752 & 1.986 & 2.026 & 1.927 & 1.661 &       \\
$17^-$ &$15^-$ & 1.376 & 1.817 & 2.069 & 2.114 & 2.010 & 1.722 &       \\
$19^-$ &$17^-$ & 1.386 & 1.875 & 2.142 & 2.190 & 2.082 & 1.777 &       \\
\hline
\end{tabular}
\end{table}

\newpage 
\parindent=0pt
%%%%%%%%%%%%%%%%%%%%%%%%%%%%%%%%%%%%%%%%%%%%%%%%%%%%%%%%%%%%%%%%%%%%%%
%%%%%%%%%%%%%%%%%%% Table 3  %%%%%%%%%%%%%%%%%%%%%%%%%%%%%%%%%%%%%%%%

\begin{table}

\caption{ $B(E1;L_i\to L_f)$ values between states with $s=1$, 
normalized to the $1_1^-\to 0_1^+$ transition.  } 

\bigskip

\begin{tabular}{r r r r r r r r }
\hline
   & $\phi_0$ & $1^{\rm o}$ & $15^{\rm o}$ & $30^{\rm o}$ & 
$45^{\rm o}$ & $60^{\rm 0}$ & $75^{\rm o}$ \\ 
$L_i^{\pi}$ & $L_f^{\pi}$ &  &  &  &  &  &  \\
 \hline
$ 1^-$ &$ 0^+$ & 1.000 & 1.000 & 1.000 & 1.000 & 1.000 & 1.000 \\
$ 2^+$ &$ 1^-$ & 1.200 & 1.227 & 1.264 & 1.269 & 1.247 & 1.215 \\
$ 3^-$ &$ 2^+$ & 1.286 & 1.358 & 1.455 & 1.469 & 1.414 & 1.328 \\
$ 4^+$ &$ 3^-$ & 1.334 & 1.467 & 1.633 & 1.657 & 1.564 & 1.413 \\
$ 5^-$ &$ 4^+$ & 1.364 & 1.570 & 1.807 & 1.842 & 1.712 & 1.489 \\
$ 6^+$ &$ 5^-$ & 1.385 & 1.670 & 1.977 & 2.022 & 1.857 & 1.562 \\
$ 7^-$ &$ 6^+$ & 1.401 & 1.768 & 2.141 & 2.196 & 2.000 & 1.634 \\
$ 8^+$ &$ 7^-$ & 1.413 & 1.864 & 2.299 & 2.364 & 2.139 & 1.705 \\
$ 9^-$ &$ 8^+$ & 1.423 & 1.957 & 2.449 & 2.523 & 2.273 & 1.777 \\
$10^+$ &$ 9^-$ & 1.431 & 2.048 & 2.592 & 2.676 & 2.403 & 1.847 \\
$11^-$ &$10^+$ & 1.437 & 2.135 & 2.727 & 2.821 & 2.527 & 1.917 \\
$12^+$ &$11^-$ & 1.443 & 2.220 & 2.856 & 2.959 & 2.646 & 1.985 \\
$13^-$ &$12^+$ & 1.448 & 2.300 & 2.979 & 3.090 & 2.760 & 2.052 \\
$14^+$ &$13^-$ & 1.452 & 2.377 & 3.095 & 3.215 & 2.870 & 2.117 \\
$15^-$ &$14^+$ & 1.456 & 2.451 & 3.206 & 3.334 & 2.974 & 2.181 \\
$16^+$ &$15^-$ & 1.460 & 2.522 & 3.311 & 3.447 & 3.075 & 2.242 \\
$17^-$ &$16^+$ & 1.463 & 2.590 & 3.411 & 3.555 & 3.171 & 2.302 \\
$18^+$ &$17^-$ & 1.466 & 2.655 & 3.507 & 3.659 & 3.263 & 2.360 \\
$19^-$ &$18^+$ & 1.469 & 2.718 & 3.598 & 3.757 & 3.351 & 2.417 \\
$20^+$ &$19^-$ & 1.471 & 2.778 & 3.685 & 3.852 & 3.436 & 2.471 \\
\hline 
\end{tabular}
\end{table}

\newpage 
\parindent=0pt
%%%%%%%%%%%%%%%%%%%%%%%%%%%%%%%%%%%%%%%%%%%%%%%%%%%%%%%%%%%%%%%%%%%%%%
%%%%%%%%%%%%%%%%%%% Table 4  %%%%%%%%%%%%%%%%%%%%%%%%%%%%%%%%%%%%%%%%

\begin{table}

\caption{ $B(E3; L_i\to L_f)$ values between states with $s=1$, normalized 
to the $3_1^-\to 0_1^+$ transition. } 

\bigskip

\begin{tabular}{r r r r r r r r }
\hline
  & $\phi_0$ & $1^{\rm o}$ & $15^{\rm o}$ & $30^{\rm o}$ & 
$45^{\rm o}$ & $60^{\rm 0}$ & $75^{\rm o}$ \\ 
$L_i^{\pi}$ & $L_f^{\pi}$ &  &  &  &  &  &  \\
 \hline
$ 3^-$ &$ 0^+$ & 1.000 & 1.000 & 1.000 & 1.000 & 1.000 & 1.000 \\  
$ 4^+$ &$ 1^-$ & 1.333 & 1.351 & 1.369 & 1.373 & 1.364 & 1.345 \\
$ 5^-$ &$ 2^+$ & 1.515 & 1.563 & 1.613 & 1.623 & 1.596 & 1.547 \\
$ 6^+$ &$ 3^-$ & 1.632 & 1.719 & 1.809 & 1.825 & 1.778 & 1.690 \\
$ 7^-$ &$ 4^+$ & 1.714 & 1.847 & 1.979 & 2.001 & 1.934 & 1.804 \\
$ 8^+$ &$ 5^-$ & 1.774 & 1.958 & 2.130 & 2.158 & 2.073 & 1.900 \\
$ 9^-$ &$ 6^+$ & 1.821 & 2.058 & 2.267 & 2.300 & 2.199 & 1.986 \\
$10^+$ &$ 7^-$ & 1.859 & 2.149 & 2.390 & 2.429 & 2.314 & 2.063 \\
$11^-$ &$ 8^+$ & 1.889 & 2.233 & 2.502 & 2.546 & 2.420 & 2.135 \\
$12^+$ &$ 9^-$ & 1.915 & 2.310 & 2.605 & 2.653 & 2.517 & 2.202 \\
$13^-$ &$10^+$ & 1.936 & 2.381 & 2.699 & 2.752 & 2.607 & 2.264 \\
$14^+$ &$11^-$ & 1.955 & 2.447 & 2.786 & 2.842 & 2.691 & 2.323 \\
$15^-$ &$12^+$ & 1.971 & 2.509 & 2.865 & 2.926 & 2.768 & 2.379 \\
$16^+$ &$13^-$ & 1.985 & 2.567 & 2.939 & 3.004 & 2.841 & 2.431 \\
$17^-$ &$14^+$ & 1.998 & 2.621 & 3.008 & 3.076 & 2.908 & 2.481 \\
$18^+$ &$15^-$ & 2.009 & 2.671 & 3.072 & 3.143 & 2.971 & 2.528 \\
$19^-$ &$16^+$ & 2.019 & 2.719 & 3.132 & 3.207 & 3.031 & 2.573 \\
$20^+$ &$17^-$ & 2.028 & 2.763 & 3.188 & 3.266 & 3.087 & 2.615 \\
\hline
%   &   &       &       &       &       &       &       \\
$ 2^+$ &$ 1^-$ & 1.800 & 1.794 & 1.797 & 1.797 & 1.793 & 1.795 \\
$ 3^-$ &$ 2^+$ & 1.333 & 1.353 & 1.386 & 1.390 & 1.370 & 1.344 \\
$ 4^+$ &$ 3^-$ & 1.273 & 1.321 & 1.383 & 1.391 & 1.356 & 1.301 \\
$ 5^-$ &$ 4^+$ & 1.259 & 1.338 & 1.429 & 1.441 & 1.393 & 1.308 \\
$ 6^+$ &$ 5^-$ & 1.257 & 1.370 & 1.486 & 1.503 & 1.442 & 1.329 \\
$ 7^-$ &$ 6^+$ & 1.258 & 1.406 & 1.545 & 1.565 & 1.495 & 1.355 \\
$ 8^+$ &$ 7^-$ & 1.261 & 1.443 & 1.602 & 1.625 & 1.547 & 1.383 \\
$ 9^-$ &$ 8^+$ & 1.264 & 1.480 & 1.656 & 1.683 & 1.598 & 1.413 \\
$10^+$ &$ 9^-$ & 1.267 & 1.516 & 1.707 & 1.736 & 1.646 & 1.442 \\
$11^-$ &$10^+$ & 1.270 & 1.550 & 1.754 & 1.786 & 1.691 & 1.471 \\
$12^+$ &$11^-$ & 1.272 & 1.582 & 1.798 & 1.833 & 1.734 & 1.498 \\
$13^-$ &$12^+$ & 1.275 & 1.612 & 1.840 & 1.876 & 1.774 & 1.526 \\
$14^+$ &$13^-$ & 1.277 & 1.641 & 1.878 & 1.917 & 1.812 & 1.552 \\
$15^-$ &$14^+$ & 1.279 & 1.669 & 1.914 & 1.955 & 1.848 & 1.577 \\
$16^+$ &$15^-$ & 1.281 & 1.695 & 1.948 & 1.991 & 1.881 & 1.601 \\
$17^-$ &$16^+$ & 1.283 & 1.719 & 1.979 & 2.025 & 1.913 & 1.624 \\
$18^+$ &$17^-$ & 1.285 & 1.742 & 2.009 & 2.056 & 1.943 & 1.646 \\
$19^-$ &$18^+$ & 1.287 & 1.764 & 2.037 & 2.086 & 1.971 & 1.667 \\
$20^+$ &$19^-$ & 1.288 & 1.785 & 2.064 & 2.114 & 1.998 & 1.688 \\
\hline 
\end{tabular}
\end{table}

\newpage 
\parindent=0pt
%%%%%%%%%%%%%%%%%%%%%%%%%%%%%%%%%%%%%%%%%%%%%%%%%%%%%%%%%%%%%%%%%%%%%%
%%%%%%%%%%%%%%%%%%% Table 5 %%%%%%%%%%%%%%%%%%%%%%%%%%%%%%%%%%%%%%%%

\begin{table}

\caption{ Experimental $B(E1; L \to L-1)$ / $B(E2; L \to L-2)$ ratios
(multiplied by $10^5$) \cite {Bizzeti} of $B(E1)$ and $B(E2)$ values 
originating 
from the same level of $^{226}$Th, compared to theoretical predictions for 
$\phi_0=45^{\rm o}$, $60^{\rm o}$, as well as to theoretical predictions 
by Bizzeti and Bizzeti-Sona \cite{Bizzeti} (labeled as BBS). As in 
Ref. \cite{Bizzeti}, asterisks indicate the values used for normalization.
}

\bigskip

\begin{tabular}{r l l l l}
\hline
$L^{\pi}$  & exp  & $45^{\rm o}$ & $60^{\rm o}$ & BBS \\ 
\hline
$ 8^+$ & 2.0 (8)     & 1.454     & 1.457     & 1.3     \\
$ 9^-$ & 1.7 (2)     & 1.649     & 1.652     & 1.3     \\
$10^+$ & 1.5 (1)$^*$ & 1.500$^*$ & 1.500$^*$ & 1.6$^*$ \\
$11^-$ & 1.7 (1)$^*$ & 1.700$^*$ & 1.700$^*$ & 1.6$^*$ \\
$12^+$ & 1.6 (1)     & 1.544     & 1.542     & 1.8     \\
$13^-$ &             & 1.747     & 1.746     & 1.8     \\
$14^+$ & 1.4 (1)     & 1.585     & 1.582     & 1.9     \\
$15^-$ & 1.7 (3)     & 1.791     & 1.789     & 2.0     \\
$16^+$ &             & 1.622     & 1.619     & 2.1     \\
$17^-$ & 1.5 (3)     & 1.831     & 1.829     & 2.1     \\
$18^+$ &             & 1.656     & 1.653     & 2.2     \\
$19^-$ & 1.7 (4)     & 1.867     & 1.865     & 2.3     \\
\hline 
\end{tabular}
\end{table}

\newpage 
\parindent=0pt
%%%%%%%%%%%%%%%%%%%%%%%%%%%%%%%%%%%%%%%%%%%%%%%%%%%%%%%%%%%%%%%%%%%%%%
%%%%%%%%%%%%%%%%%%% Table 6 %%%%%%%%%%%%%%%%%%%%%%%%%%%%%%%%%%%%%%%%

\begin{table}

\caption{Experimental $B(E1; L\to L+1)$ / $B(E1; L\to L-1)$ ratios \cite{Woll}
of $B(E1)$ values originating from the same level of $^{226}$Ra, compared to 
theoretical predictions for $\phi_0=45^{\rm o}$, $56^{\rm o}$, as well as 
to three different theoretical predictions from Ref. \cite{Raduta}, 
labelled as R-h, R-I, R-II. See section 4 for further discussion.  
}

\bigskip

\begin{tabular}{r r r r r r r}
\hline
$L^{\pi}$  & exp & $45^{\rm o}$ & $56^{\rm o}$ & R-h & R-I & R-II \\
\hline
$ 1^-$ & 1.85$\pm$1.20 & 2.116 & 2.092 & 1.84 & 1.98 & 1.85 \\
$ 3^-$ & 0.87$\pm$0.35 & 1.451 & 1.433 & 1.12 & 1.31 & 0.95 \\
$ 5^-$ &               & 1.297 & 1.288 &      &      &      \\
$ 7^-$ & 1.79$\pm$1.59 & 1.220 & 1.215 & 0.86 & 1.12 & 0.99 \\
$ 9^-$ & 1.27$\pm$0.68 & 1.172 & 1.170 & 0.83 & 1.10 & 1.13 \\
$11^-$ & 1.12$\pm$0.79 & 1.140 & 1.139 & 0.83 & 1.10 & 1.26 \\
$13^-$ & 1.06$\pm$0.68 & 1.117 & 1.117 & 0.85 & 1.11 & 1.35 \\
\hline 
\end{tabular}
\end{table}

\newpage 
\parindent=0pt
%%%%%%%%%%%%%%%%%%%%%%%%%%%%%%%%%%%%%%%%%%%%%%%%%%%%%%%%%%%%%%%%%%%%%%
%%%%%%%%%%%%%%%%%%% Table 7 %%%%%%%%%%%%%%%%%%%%%%%%%%%%%%%%%%%%%%%%

\begin{table}

\caption{Parameter values $\beta_{0,max}$ where the first derivative of the 
energy ratios $R(L)=E(L)/E(2)$ for the ground state band and the associated 
negative parity band ($s=1$) of the Davidson potentials of Eq. (\ref{eq:e15})
has a maximum, while the second derivative vanishes, together with the $R(L)$ 
ratios obtained at 
these values (labeled by ``var'') and the corresponding ratios of the 
present model (labeled by ``oct''), for several values of the angular 
momentum $L$. In all cases $\phi_0=45^{\rm o}$ has been used. 
See section 5 for further discussion. 
}

\bigskip

\begin{tabular}{r r r r}
\hline
$L^{\pi}$ & $\beta_{0,max}$ & $R_L$ & $R_L$ \\
  &               & var   & oct \\
 \hline
$ 1^-$ & 1.200 & 0.347 & 0.346 \\
$ 2^+$ &       & 1.000 & 1.000 \\
$ 3^-$ & 1.283 & 1.909 & 1.912 \\
$ 4^+$ & 1.329 & 3.030 & 3.039 \\
$ 5^-$ & 1.374 & 4.333 & 4.351 \\
$ 6^+$ & 1.419 & 5.797 & 5.829 \\
$ 7^-$ & 1.461 & 7.407 & 7.459 \\
$ 8^+$ & 1.502 & 9.154 & 9.233 \\
$ 9^-$ & 1.541 &11.032 &11.144 \\
$10^+$ & 1.579 &13.034 &13.187  \\
\hline 
\end{tabular}
\end{table}

\newpage

%%%%% Fig. 1 %%%%%%%%%%%%%%%%%%%%%%%%%%%%%%%%%%%%%%%%%%%%%%%%%%%%%%%%%%%%%

\begin{figure}[ht]
\rotatebox{270}{\includegraphics[height=80mm]{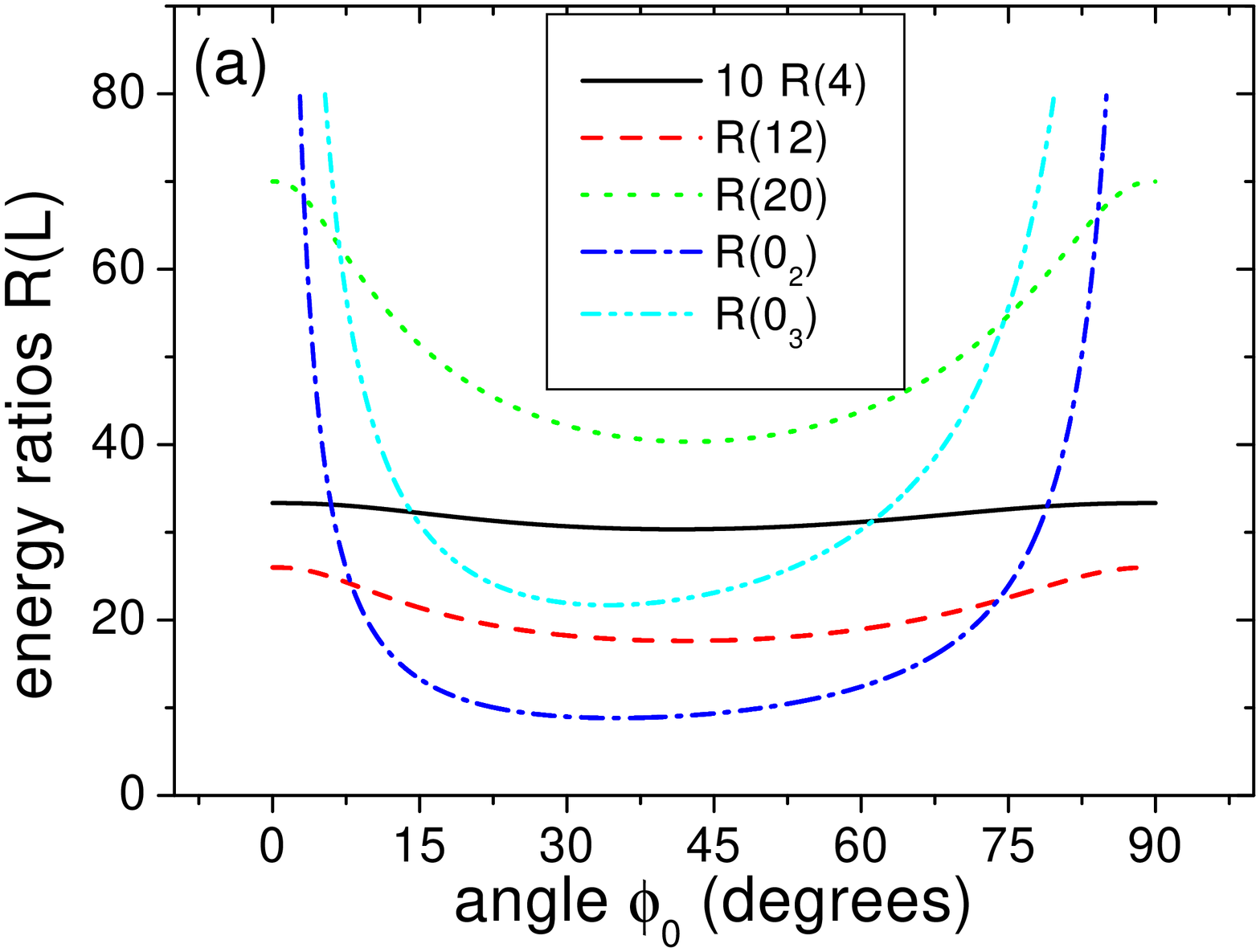}} 
\rotatebox{270}{\includegraphics[height=80mm]{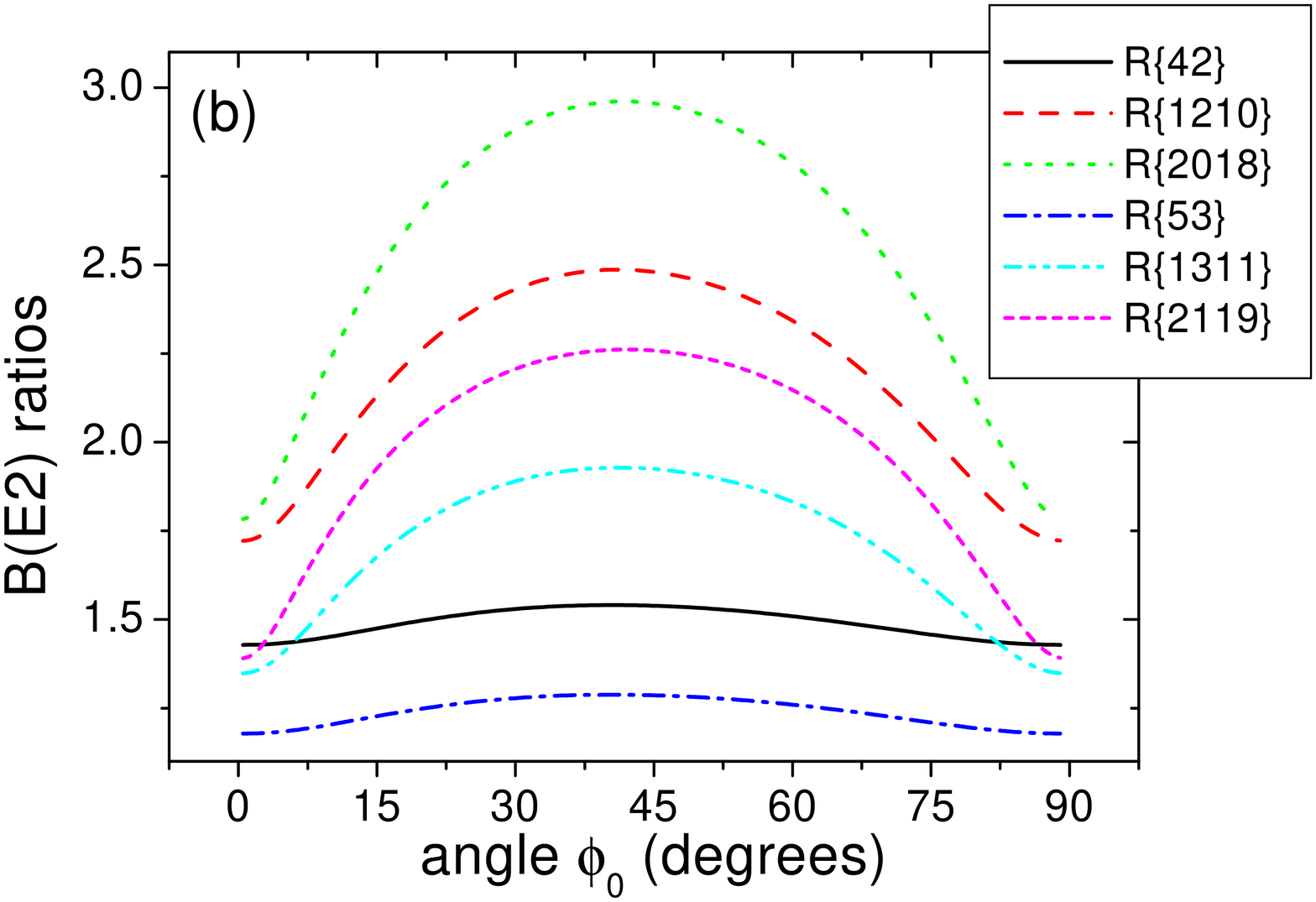}}
\rotatebox{270}{\includegraphics[height=80mm]{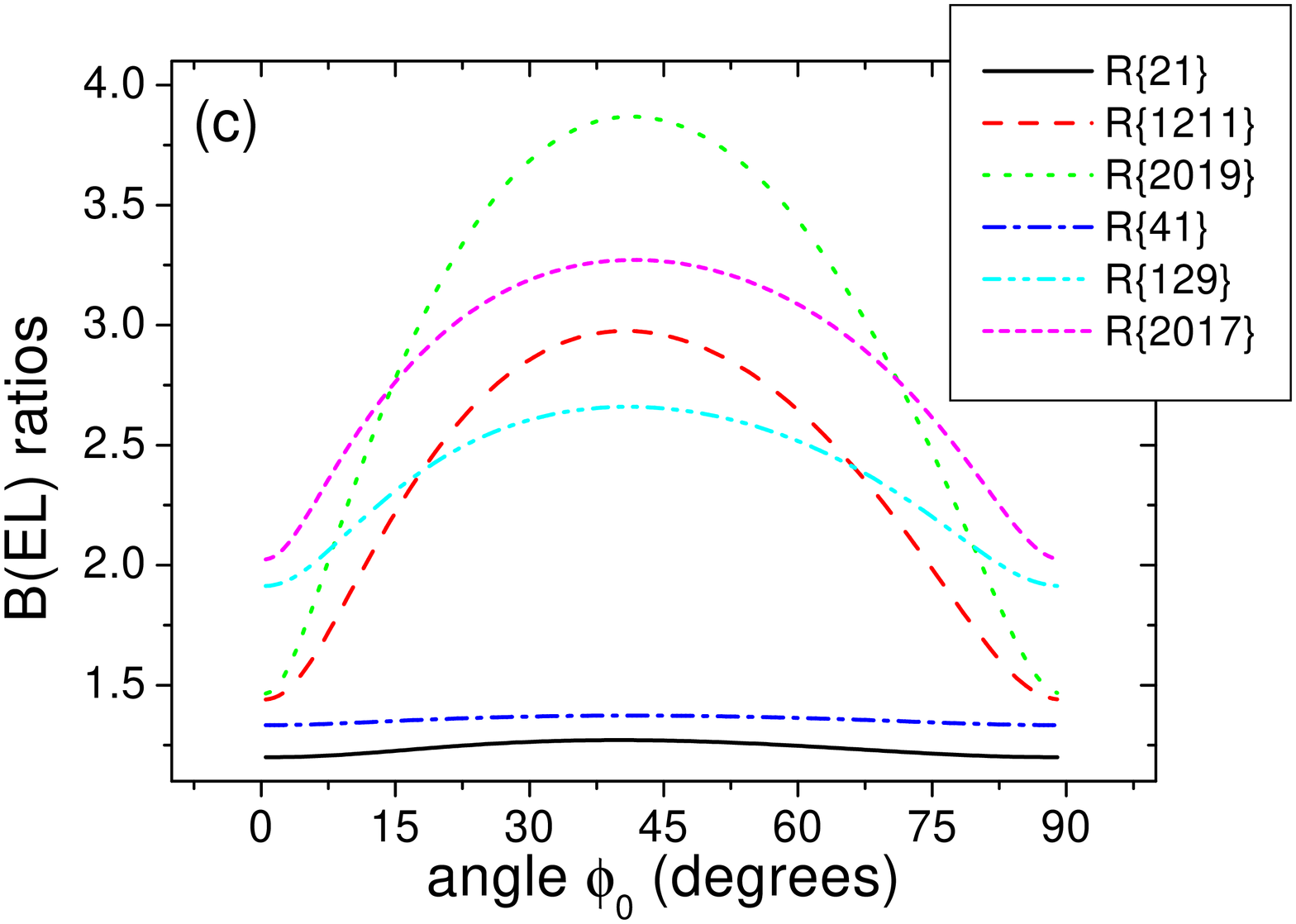}}
\caption{ (color online)
(a) Energy ratios $R(L)={E(L)\over E(2_1^+)}$ as a function of the 
angle $\phi_0$. The minima appear as follows: $R(4)$ at $41.19^{\rm o}$, 
$R(12)$ at $42.03^{\rm o}$, $R(20)$ at $42.61^{\rm o}$, $R(0_2)$ at 
$34.99^{\rm o}$, $R(0_3)$ at $34.13^{\rm o}$. 
(b) Same for the $B(E2)$ ratios 
$R\{42\}  ={B(E2;4^+\to 2^+) \over  B(E2;2^+\to 0^+)}$ ($40.5^{\rm o}$), 
$R\{1210\}={B(E2;12^+\to 10^+)\over B(E2;2^+\to 0^+)}$ ($41.0^{\rm o}$),
$R\{2018\}={B(E2;20^+\to 18^+)\over B(E2;2^+\to 0^+)}$ ($42.0^{\rm o}$),  
$R\{53\}  ={B(E2;5^-\to 3^-)\over   B(E2;3^-\to 1^-)}$ ($40.5^{\rm o}$), 
$R\{1311\}={B(E2;13^-\to 11^-)\over B(E2;3^-\to 1^-)}$ ($41.5^{\rm o}$),
$R\{2119\}={B(E2;21^-\to 19^-)\over B(E2;3^-\to 1^-)}$ ($42.0^{\rm o}$). 
After each ratio, the position of the maximum (with accuracy $\pm 0.5^{\rm o}$)
appears in parentheses. 
(c) Same for the $B(E1)$ ratios 
$R\{21\}  ={B(E1;2^+\to 1^-) \over  B(E1;1^-\to 0^+)}$ ($40.0^{\rm o}$), 
$R\{1211\}={B(E1;12^+\to 11^-)\over B(E1;1^-\to 0^+)}$ ($40.5^{\rm o}$),
$R\{2019\}={B(E1;20^+\to 19^-)\over B(E1;1^-\to 0^+)}$ ($41.5^{\rm o}$),
and for the $B(E3)$ ratios
$R\{41\}  ={B(E3;4^+\to 1^-)\over   B(E3;3^-\to 0^+)}$ ($41.0^{\rm o}$), 
$R\{129\}={B(E3;12^+\to 9^-)\over  B(E3;3^-\to 0^+)}$ ($41.0^{\rm o}$),
$R\{2017\}={B(E3;20^+\to 17^-)\over B(E3;3^-\to 0^+)}$ ($41.5^{\rm o}$). 
Again, each ratio is followed by the position of the maximun (with accuracy 
$\pm 0.5^{\rm o}$) in parentheses.}
\end{figure}

%%%%%%%%%%%%%%% Fig. 2 %%%%%%%%%%%%%%%%%%%%%%%%%%%%%%%%%%%%%%%%%%%%%%%%%%

\begin{figure}[ht]
\rotatebox{270}{\includegraphics[height=80mm]{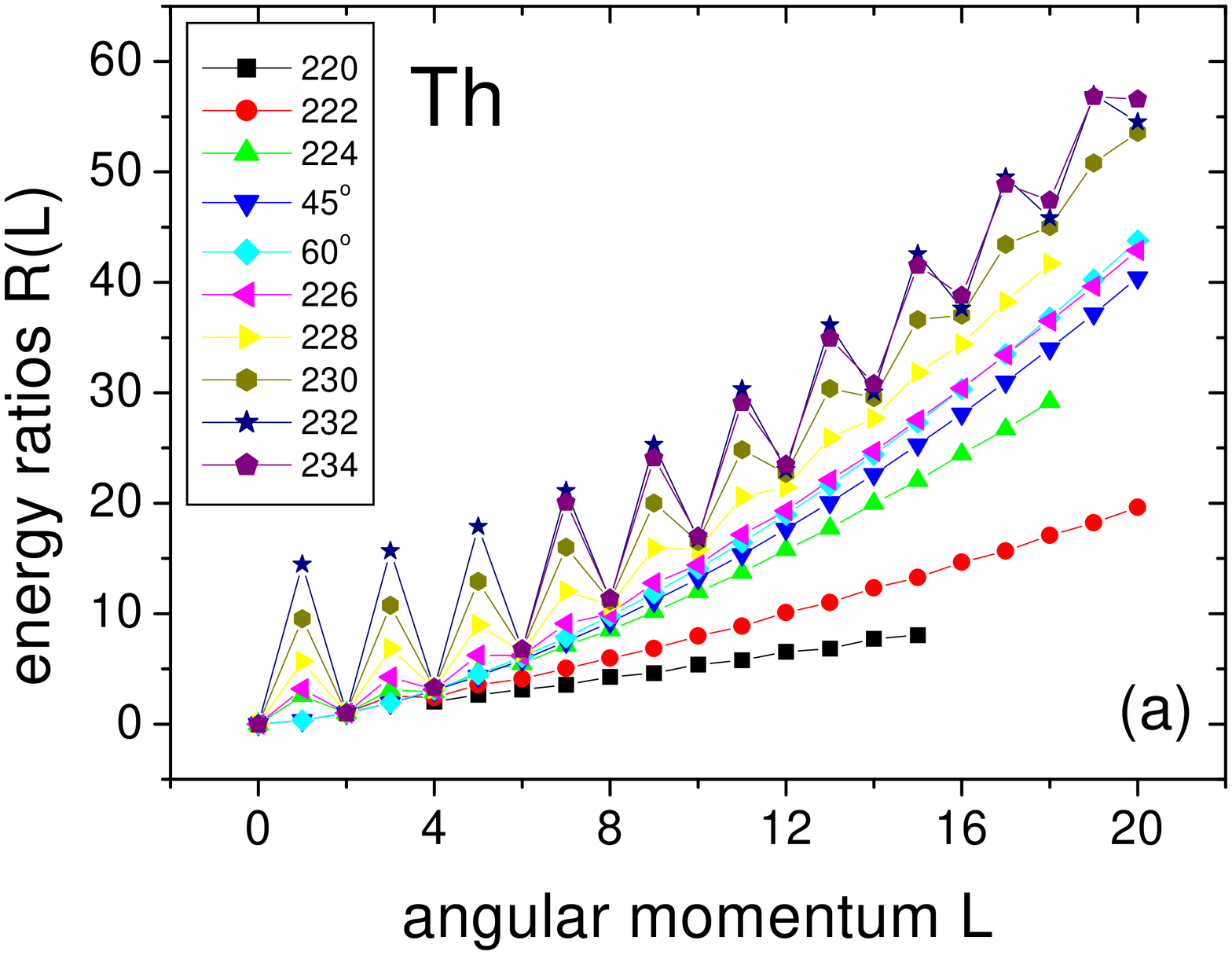}} 
\rotatebox{270}{\includegraphics[height=80mm]{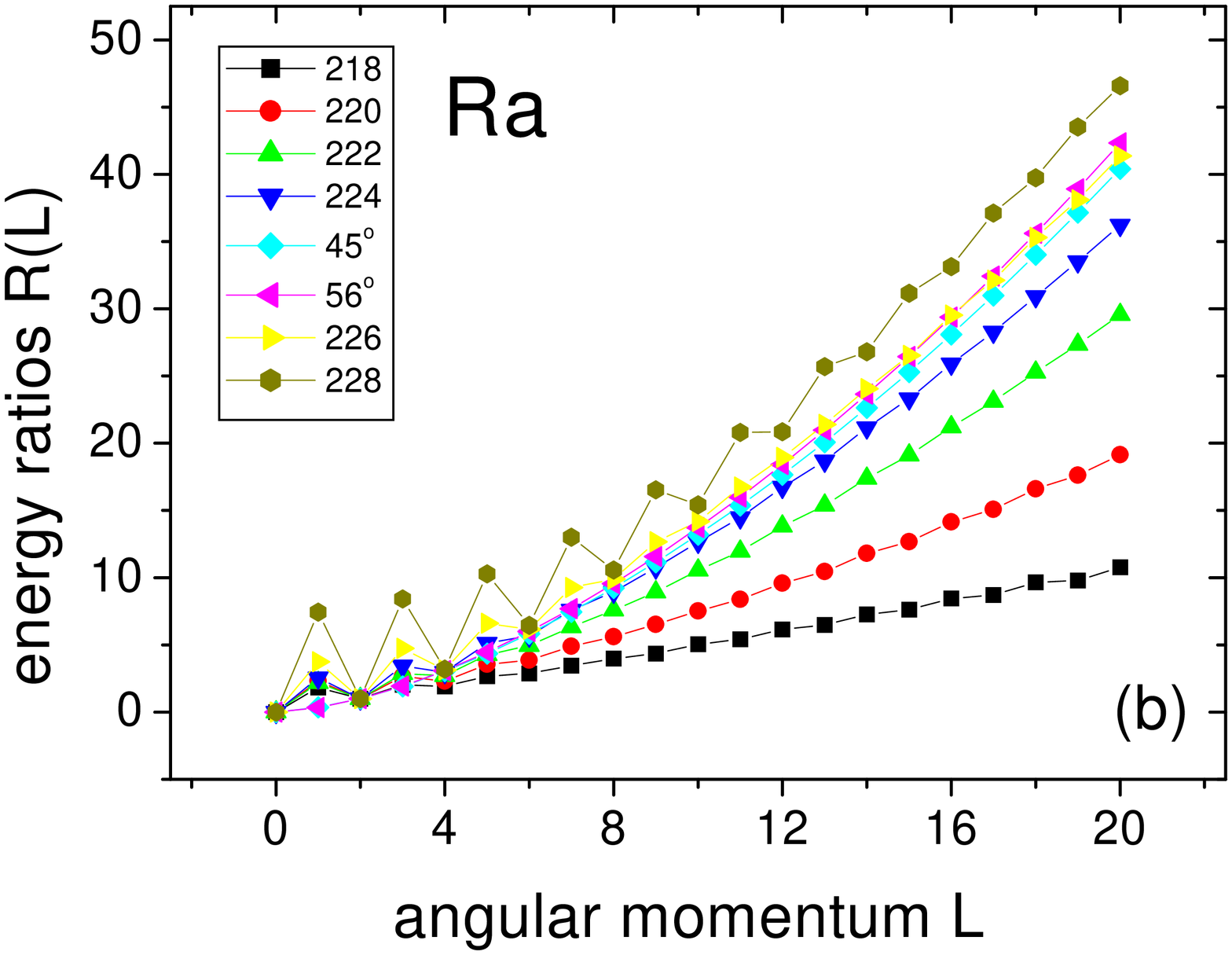}}
\caption{ (color online)
(a) Experimental energy ratios $R(L)=E(L)/E(2_1^+)$ for 
$^{220}$Th \cite{Th220}, 
$^{222}$Th \cite{Th222}, $^{224}$Th \cite{Th224}, $^{226}$Th \cite{Th226}, 
$^{228}$Th \cite{Th228}, $^{230}$Th \cite{Cocks2}, $^{232}$Th 
\cite{Cocks2,Th232}, 
and $^{234}$Th \cite{Cocks2}, compared to theoretical predictions for 
$\phi=45^{\rm o}$ and $\phi=60^{\rm o}$.
(b) Same for $^{218}$Ra \cite{Ra218,Schulz}, 
$^{220}$Ra \cite{Th220}, $^{222}$Ra \cite{Cocks2,Cocks1}, 
$^{224}$Ra \cite{Cocks2,Cocks1}, $^{226}$Ra \cite{Cocks2,Cocks1}, and 
$^{228}$Ra \cite{Cocks2}, compared to theoretical predictions for 
$\phi=45^{\rm o}$ and $\phi=56^{\rm o}$. }
\end{figure}

%%%%%%%%%%%%%%%%% Fig. 3 %%%%%%%%%%%%%%%%%%%%%%%%%%%%%%%%%%%%%%%%%%%%%%%%

\begin{figure}[ht]
\rotatebox{270}{\includegraphics[height=80mm]{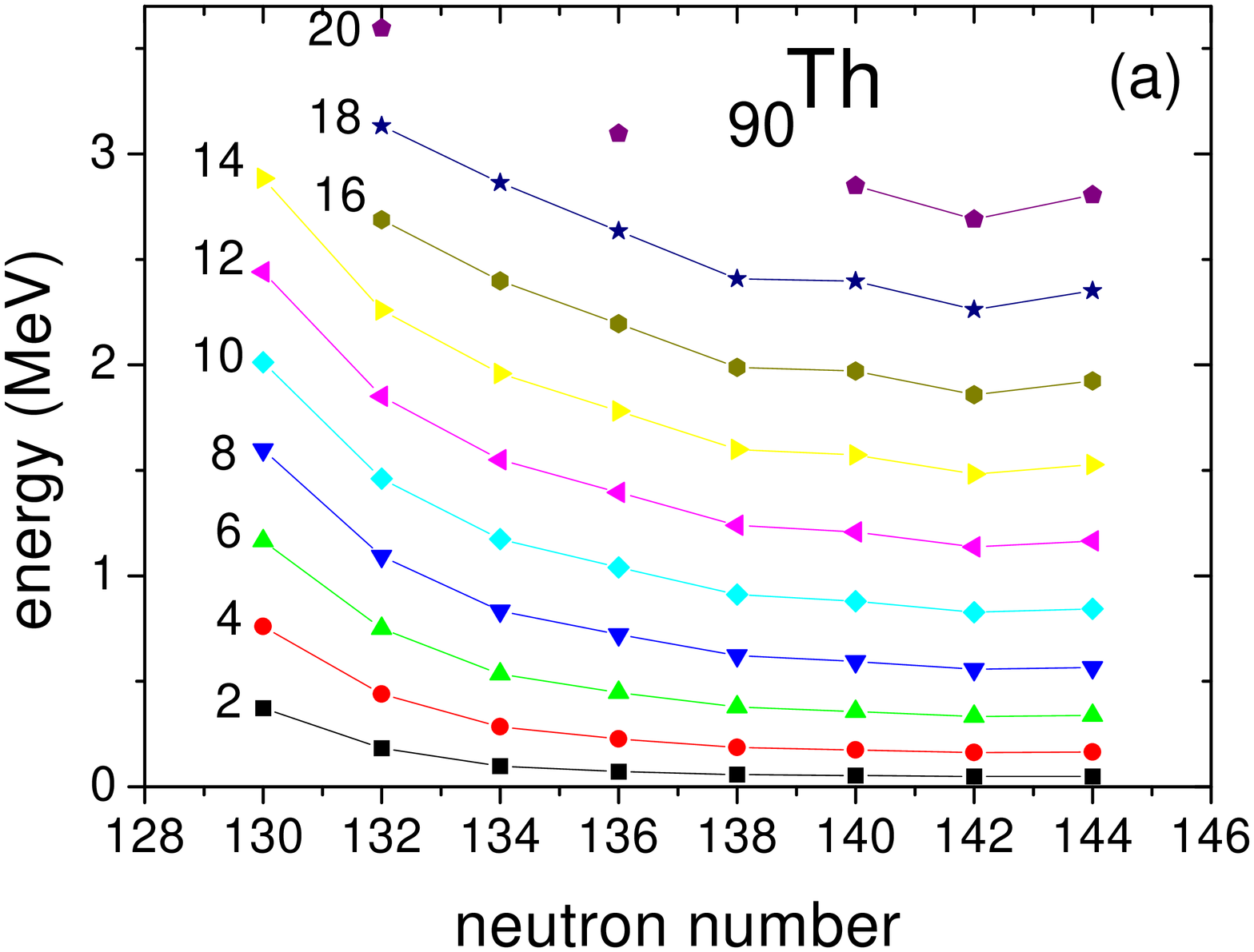}} 
\rotatebox{270}{\includegraphics[height=80mm]{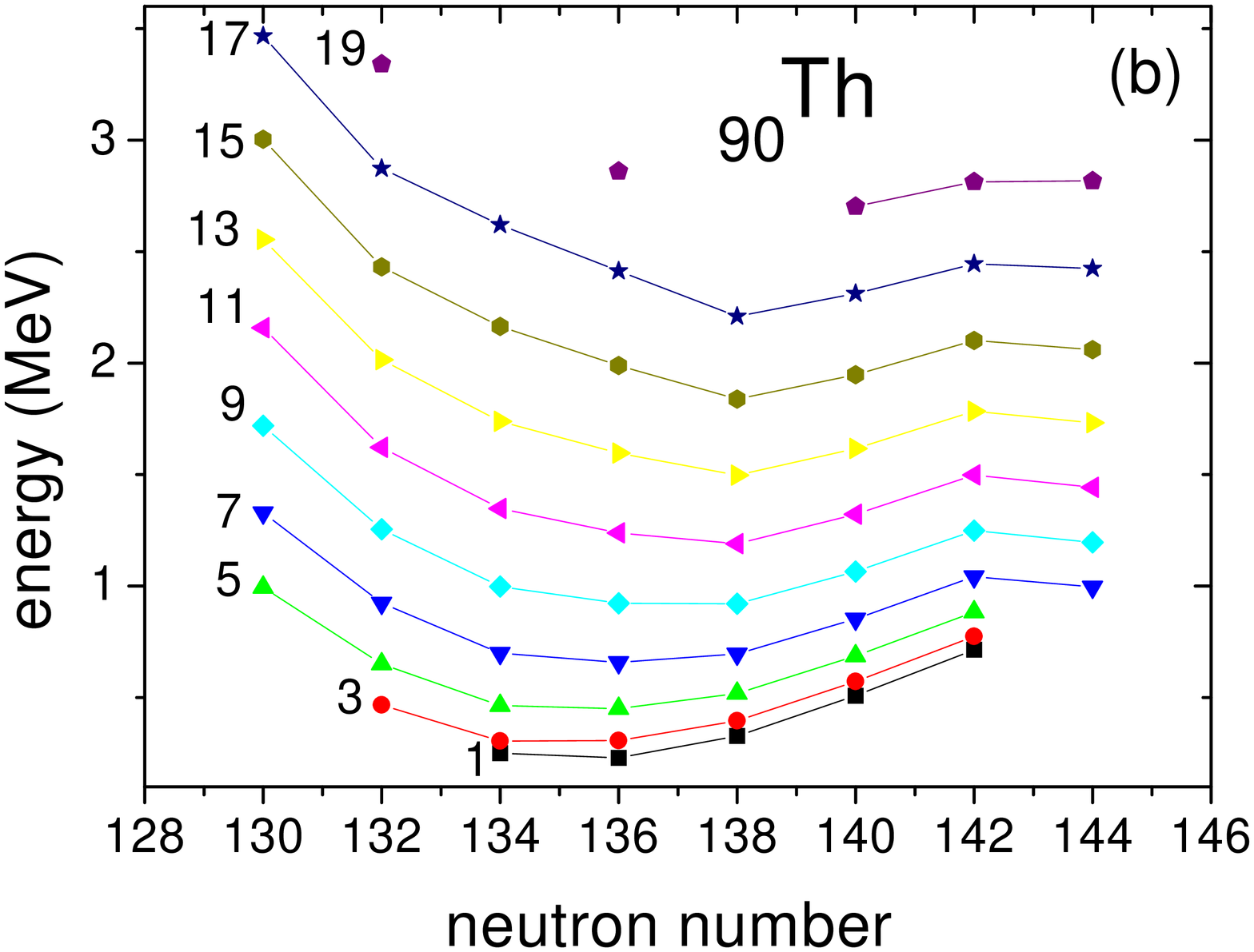}}
\rotatebox{270}{\includegraphics[height=80mm]{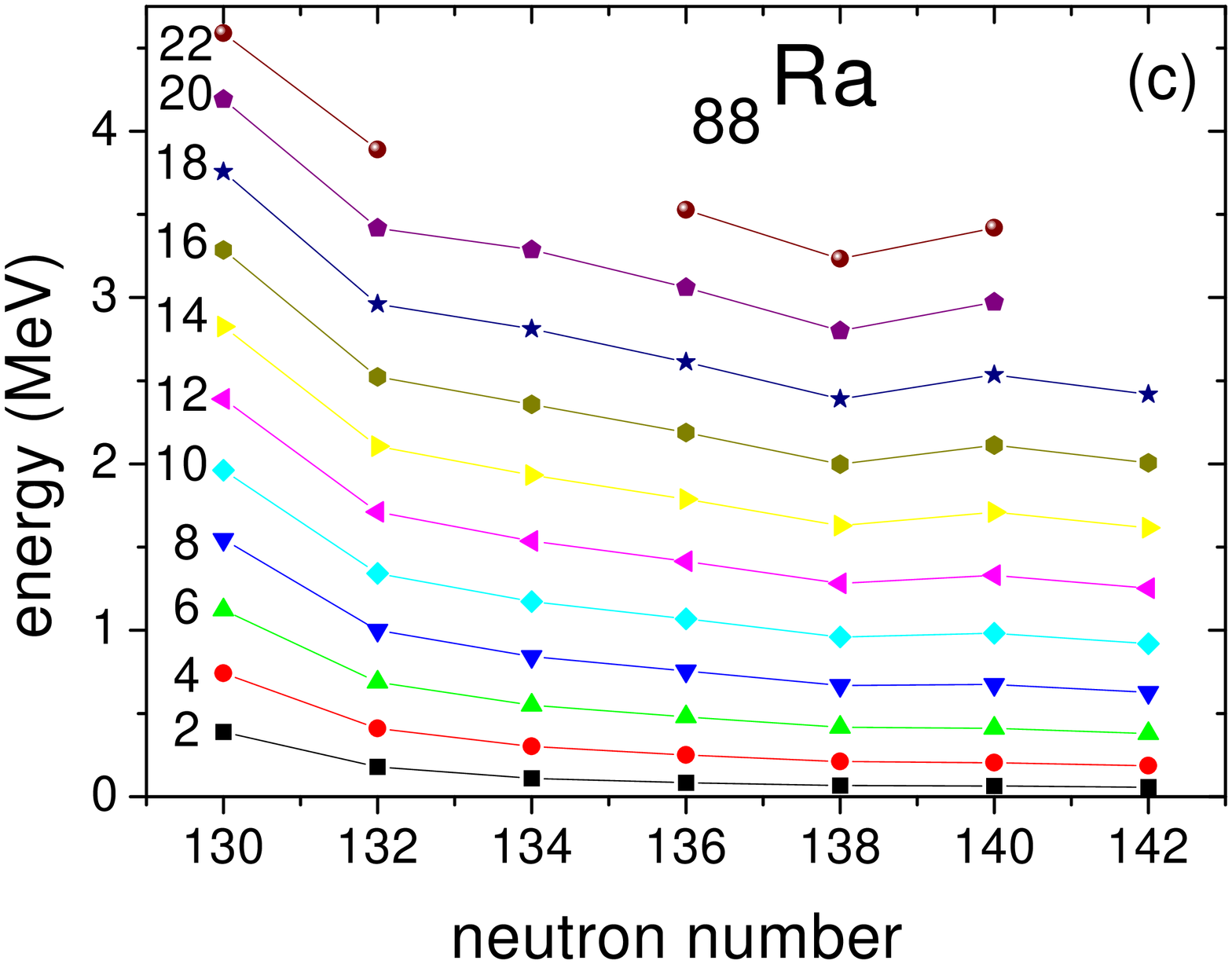}}
\rotatebox{270}{\includegraphics[height=80mm]{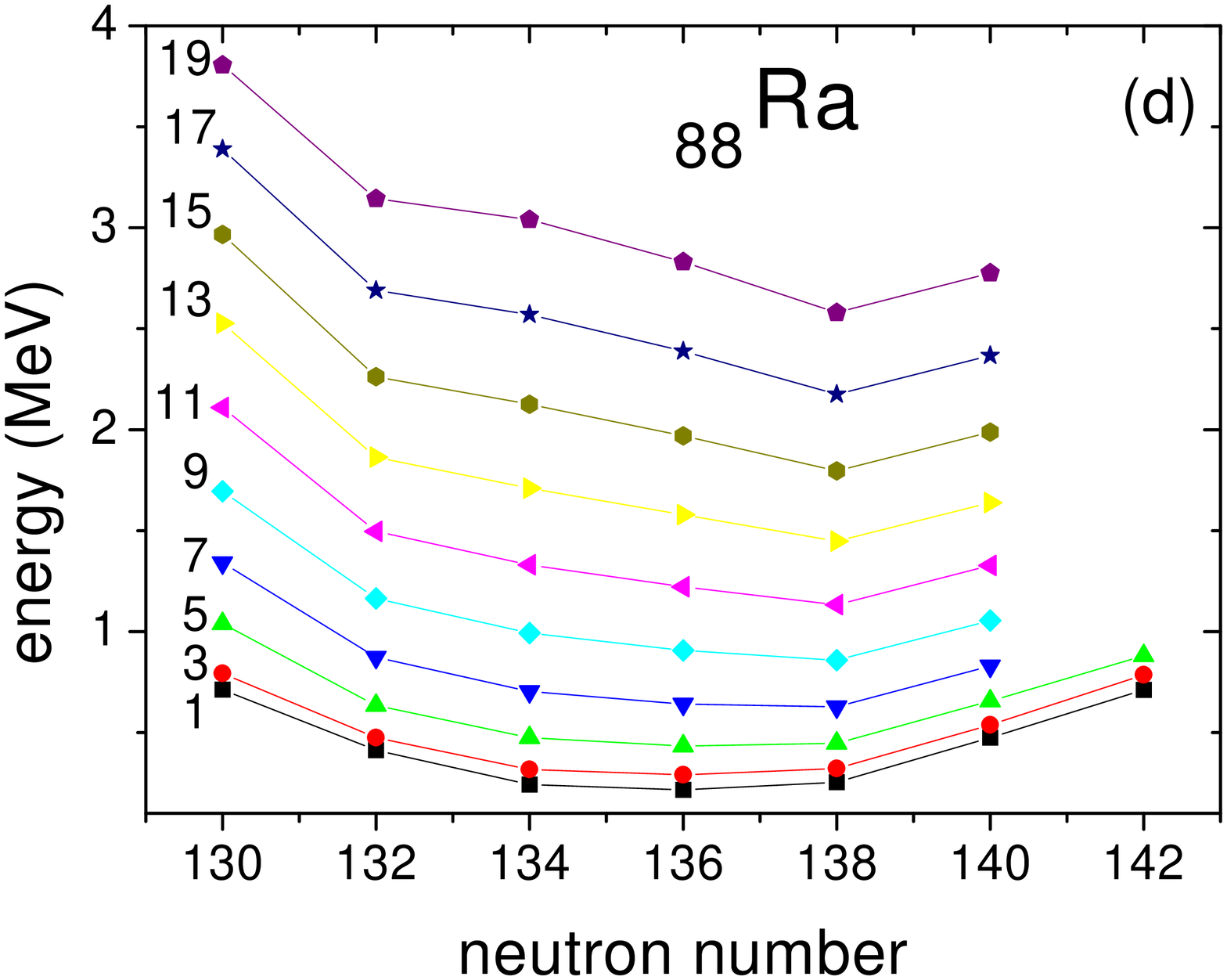}}
\caption{ (color online)
(a) Experimental energy levels of the ground state bands of 
$^{220}$Th \cite{Th220}, 
$^{222}$Th \cite{Th222}, $^{224}$Th \cite{Th224}, $^{226}$Th \cite{Th226}, 
$^{228}$Th \cite{Th228}, $^{230}$Th \cite{Cocks2}, $^{232}$Th 
\cite{Cocks2,Th232}, 
and $^{234}$Th \cite{Cocks2}, as a function of the neutron number. 
(b) Same as (a), but for the associated  negative parity bands. 
(c) Same as (a), but for $^{218}$Ra \cite{Ra218,Schulz}, 
$^{220}$Ra \cite{Th220}, $^{222}$Ra \cite{Cocks2,Cocks1}, 
$^{224}$Ra \cite{Cocks2,Cocks1}, $^{226}$Ra \cite{Cocks2,Cocks1},
$^{228}$Ra \cite{Cocks2}, and $^{230}$Ra \cite{Cocks2}.
(d) Same as (c), but for the associated negative parity bands. }
\end{figure}

%%%%%%%%%%%%%%%% Fig. 4 %%%%%%%%%%%%%%%%%%%%%%%%%%%%%%%%%%%%%%%%%%%%%%%%%%%
\begin{figure}[ht]
\rotatebox{270}{\includegraphics[height=80mm]{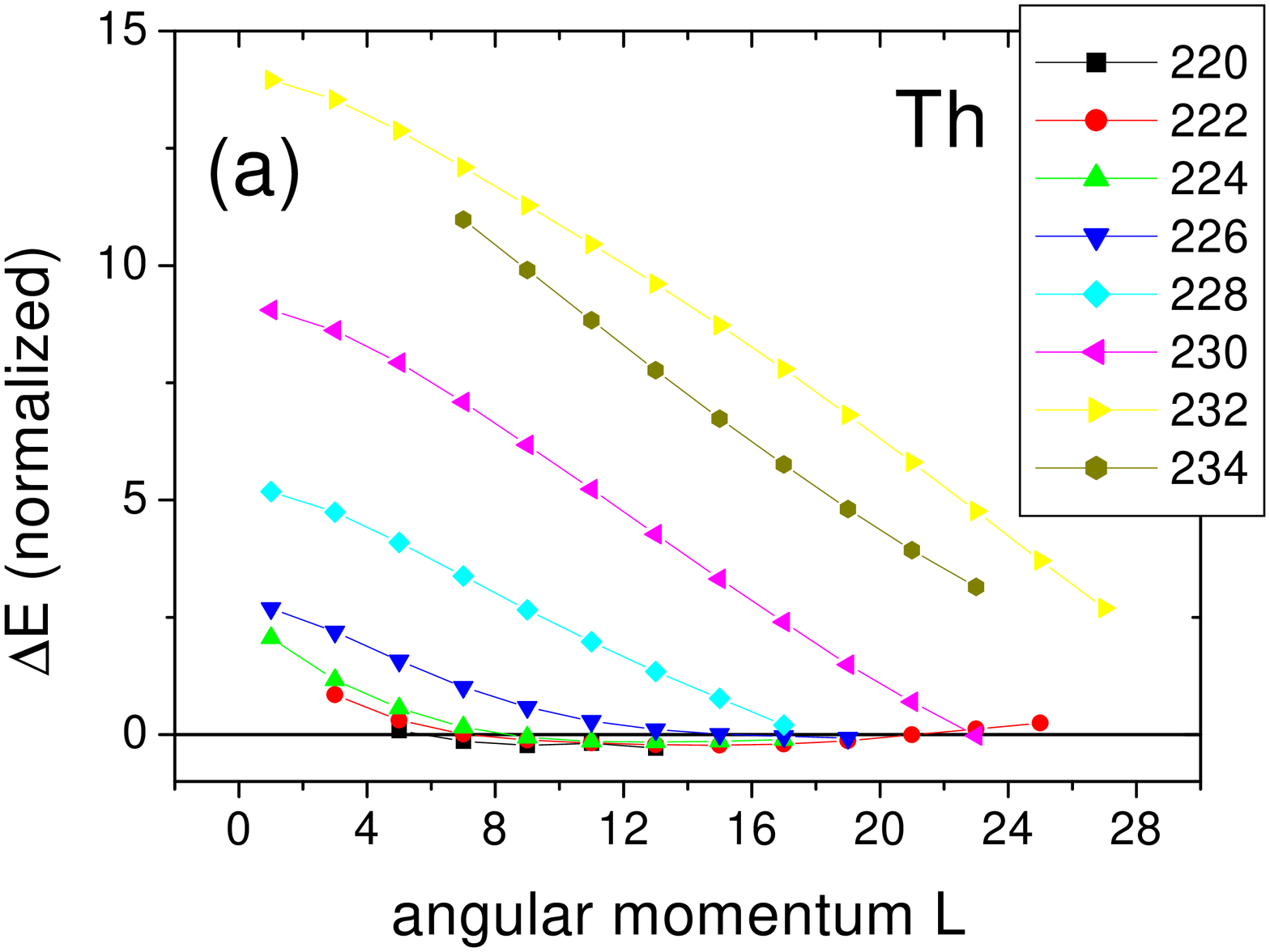}} 
\rotatebox{270}{\includegraphics[height=80mm]{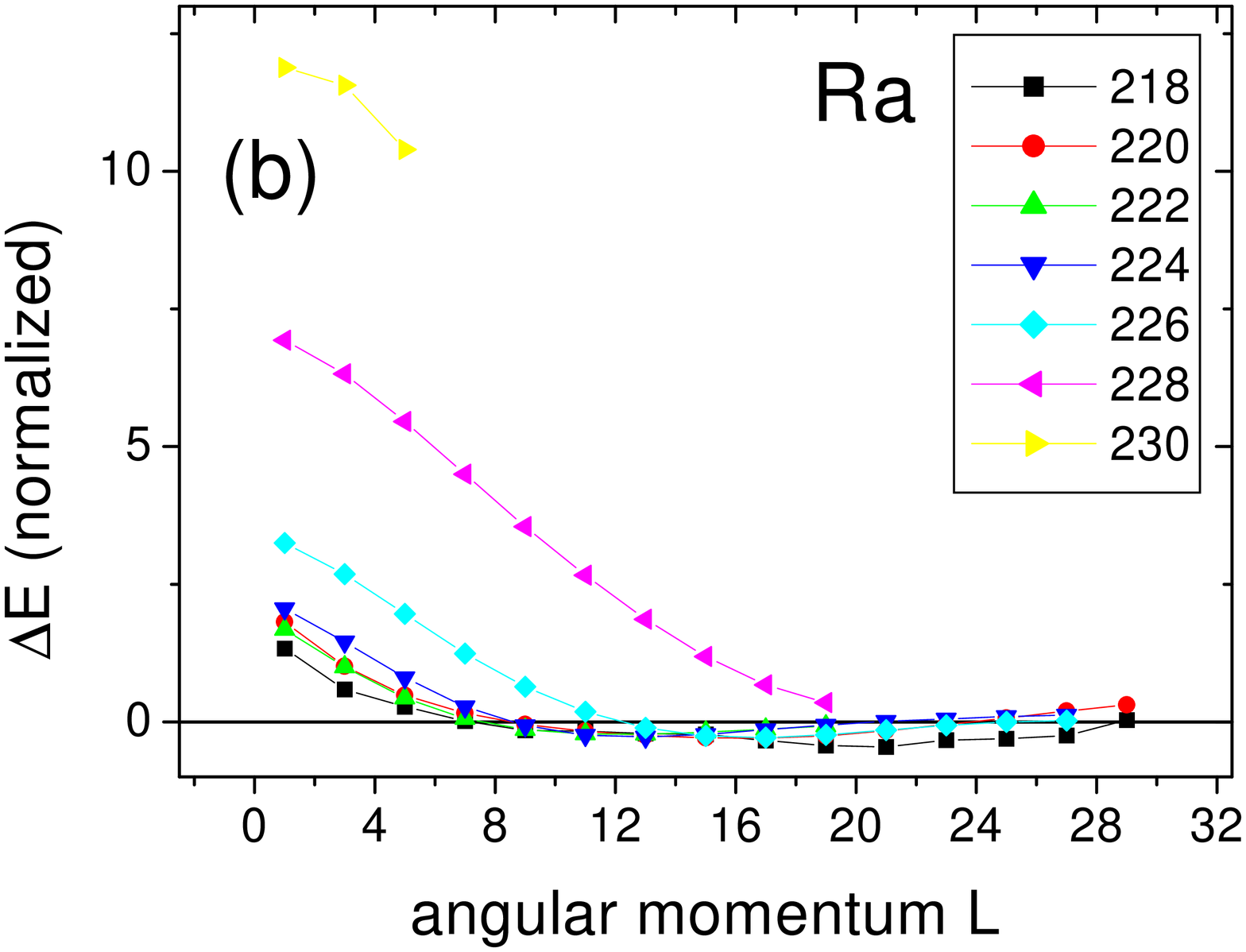}}
\caption{ (color online)
(a) Experimental energy staggering $\Delta E$ (Eq. (\ref{eq:e51})),
normalized to the $2_1^+$ state of each nucleus, for $^{220}$Th \cite{Th220}, 
$^{222}$Th \cite{Th222}, $^{224}$Th \cite{Th224}, $^{226}$Th \cite{Th226}, 
$^{228}$Th \cite{Th228}, $^{230}$Th \cite{Cocks2}, $^{232}$Th 
\cite{Cocks2,Th232}, 
and $^{234}$Th \cite{Cocks2}, as a function of the angular momentum. 
(b) Same as (a), but for $^{218}$Ra \cite{Ra218,Schulz}, 
$^{220}$Ra \cite{Th220}, $^{222}$Ra \cite{Cocks2,Cocks1}, 
$^{224}$Ra \cite{Cocks2,Cocks1}, $^{226}$Ra \cite{Cocks2,Cocks1},
$^{228}$Ra \cite{Cocks2}, and $^{230}$Ra \cite{Cocks2}.}
\end{figure}

%%%%%%%%%%%%%%%%%%%%% Fig. 5 %%%%%%%%%%%%%%%%%%%%%%%%%%%%%%%%%%%%%%%%%%%

\begin{figure}[ht]
\rotatebox{270}{\includegraphics[height=80mm]{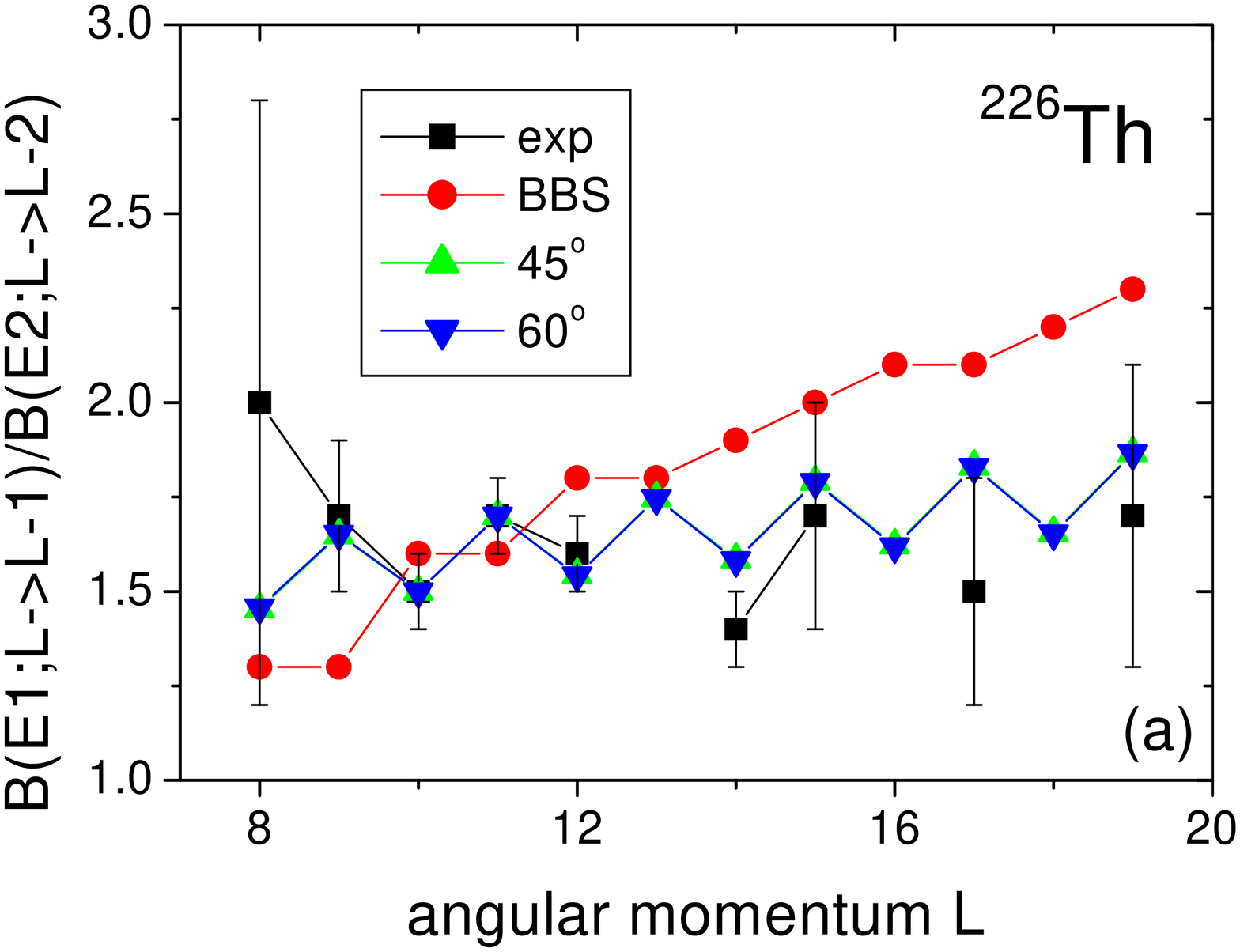}} 
\rotatebox{270}{\includegraphics[height=80mm]{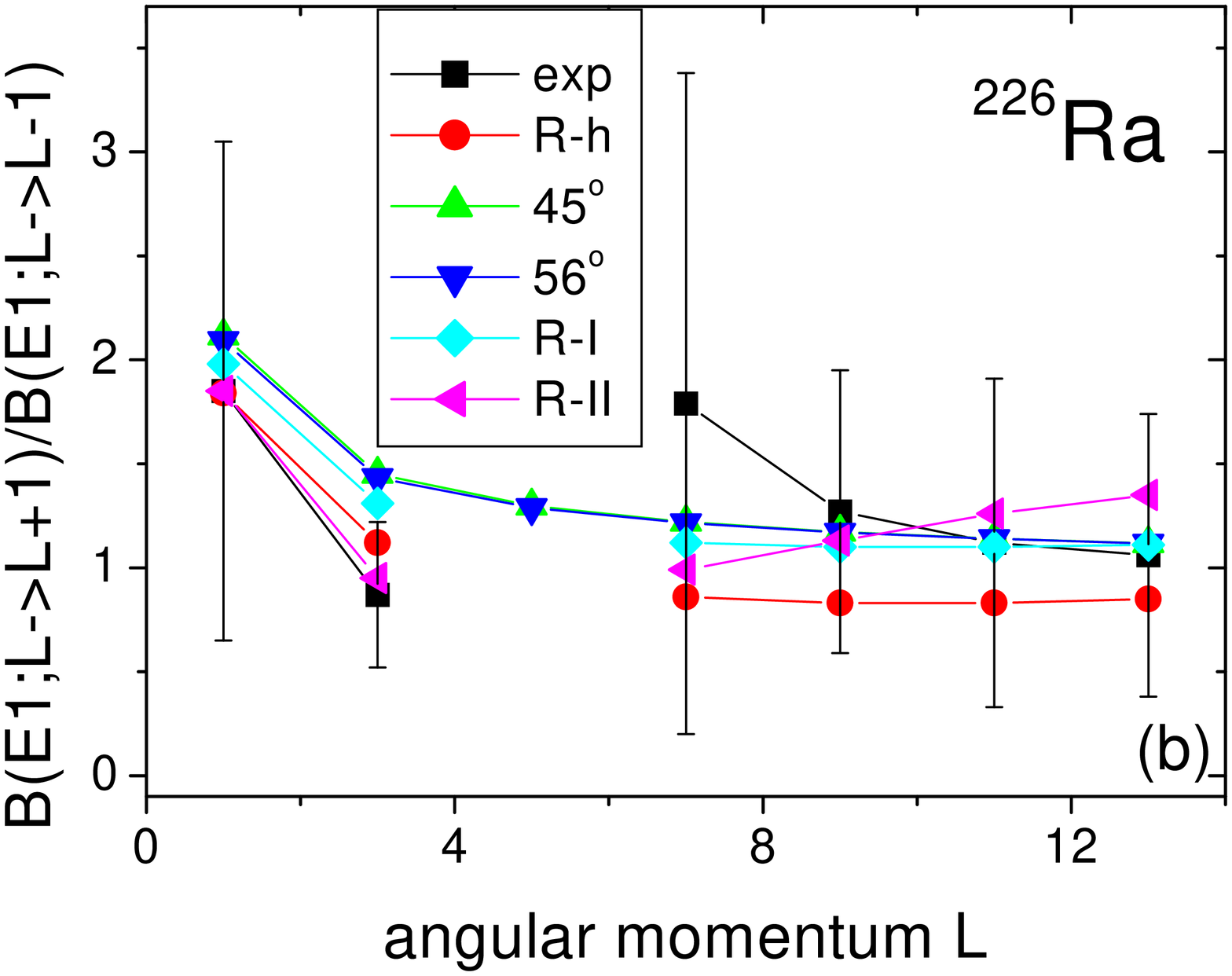}}
\caption{ (color online) 
(a) Experimental $B(E1; L \to L-1)$ / $B(E2; L \to L-2)$ ratios
(multiplied by $10^5$) \cite {Bizzeti} of $B(E1)$ and $B(E2)$
values originating 
from the same level of $^{226}$Th, compared to theoretical predictions 
by Bizzeti and Bizzeti-Sona \cite{Bizzeti} (labeled as BBS), as well as to 
predictions of the present work for $\phi_0=45^{\rm o}$, $60^{\rm o}$. 
The ratios corresponding to $L=10$ and 11 have been used for normalization, 
as in Ref. \cite{Bizzeti}. 
(b) Experimental $B(E1; L\to L+1)$ / $B(E1; L\to L-1)$ ratios \cite{Woll}
of $B(E1)$ values originating from the same level of $^{226}$Ra, compared 
to three different theoretical predictions from Ref. \cite{Raduta} (labeled 
as R-h, R-I, R-II), as well as to theoretical predictions of the present work 
for $\phi_0=45^{\rm o}$, $56^{\rm o}$. See section 4 for further discussion. }
\end{figure}

%%%%%%%%%%%%%%%%%%%%%%%% Fig. 6 %%%%%%%%%%%%%%%%%%%%%%%%%%%%%%%%%%%%%%%%%%%

\begin{figure}[ht]
\rotatebox{270}{\includegraphics[height=80mm]{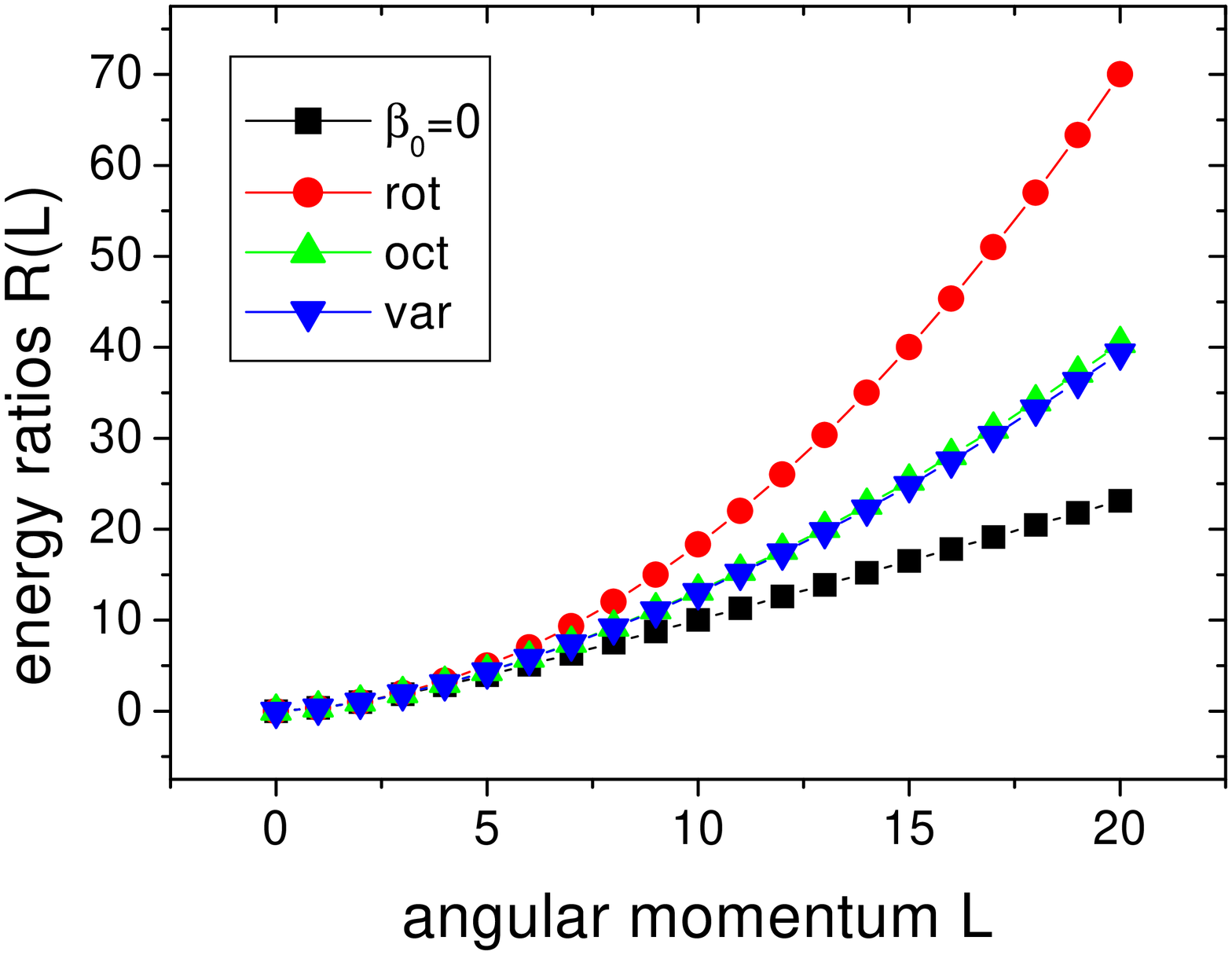}} 
\caption{ (color online) 
Energy ratios $R(L)=E(L)/E(2)$ for the ground state band and the associated 
negative parity band ($s=1$) of the Davidson potentials of Eq. (\ref{eq:e15}), 
selected through the variational procedure of section 5 (labeled by ``var''), 
compared to the ratios provided by the present work (labeled as ``oct''), 
as a function of angular momentum. The limiting cases corresponding to 
Davidson potentials with $\beta_0=0$ (labeled as ``$\beta_0=0$'') and 
$\beta_0\to \infty$ (labeled as ``rot'') are also shown for comparison. In all
cases, $\phi_0=45^{\rm o}$ has been used. See section 5 for further 
discussion.}
\end{figure}

\end{document}